\documentclass[twocolumn]{aastex631}
\usepackage{amsmath}
\usepackage{bm}
\usepackage{graphicx}
\usepackage{float}      
\usepackage{booktabs}
\usepackage{tabularx}
\usepackage{hyperref}

\definecolor{ao(english)}{rgb}{0.0, 0.5, 0.0}
\definecolor{uf(english)}{rgb}{0.0, 0.0, 0.5}
\definecolor{dy(english)}{rgb}{0.8, 0.3, 0.1}

\begin{document}
\title{1D Vlasov Simulations of QED Cascades Over Pulsar Polar Caps}
\author{Dingyi Ye}
\affiliation{Physics Department and McDonnell Center for the Space Sciences, Washington University in St. Louis; MO, 63130, USA}
\author{Alexander Y. Chen}
\affiliation{Physics Department and McDonnell Center for the Space Sciences, Washington University in St. Louis; MO, 63130, USA}

\begin{abstract}
Recent developments in the study of pulsar radio emission revealed that the microphysics of quantum electrodynamic (QED) pair cascades at pulsar polar caps may be responsible for generating the observed coherent radio waves. However, modeling the pair cascades in the polar cap region poses significant challenges, particularly under conditions of high plasma multiplicity. Traditional Particle-in-Cell (PIC) methods often face rapidly increasing computational costs as the multiplicity grows exponentially. To address this issue, we present a new simulation code using the Vlasov method, which efficiently simulates the evolution of charged particle distribution functions in phase space without a proportional increase in computational expense at high multiplicities. We apply this code to study $e^\pm$ pair cascades in 1D, incorporating key physical processes such as curvature radiation, radiative cooling, and magnetic pair production. We study both the Ruderman-Sutherland (RS) and the Space-charge-limited Flow (SCLF) regimes, and find quasiperiodic gap formation and pair production bursts in both cases. These features produce strong electric field oscillations, potentially enabling coherent low-frequency radio emission. We construct a unified analytic model that describes the key features of the polar cap cascade, which can be used to estimate the return current heating rate that can be used to inform X-ray hotspot models. Spectral analysis shows that a significant amount of energy is carried in superluminal modes -- collective excitations that could connect to observed radio features. Our results align with previous PIC studies while offering enhanced fidelity in both dense and rarefied regions.
\end{abstract}

\section{Introduction}

Pulsars are believed to be rotating neutron stars that power their pulsed emission using their rotation energy, leading to gradual spin-down over their lifetime \citep{Lorimer2005}. \citet{GoldreichJulian1969} showed that the rotation of the neutron star necessitates a corotating plasma, with a characteristic charge density \( \rho_{\rm GJ} = -\mathbf{\Omega}\cdot\mathbf{B}/2\pi c \). This plasma is believed to be generated via electromagnetic pair cascades, most likely near the magnetic polar caps \citep[e.g.,][]{Ruderman1975, AronsScharlemann1979, Hibschman2001}. In addition to providing the background charge density $\rho_\mathrm{GJ}$, this plasma also supports the magnetospheric current density $\bm{j}$ required by the global magnetic field configuration~\citep[see e.g.,][]{1973ApJ...180..207M, contopoulos_axisymmetric_1999}. When there is enough plasma to support both $\rho_\mathrm{GJ}$ and the magnetospheric current, the magnetosphere is \emph{Force-free}~\citep[e.g.,][]{Gruzinov1999, Spitkovsky2006}. The structure of the force-free magnetosphere has been studied extensively~\citep[e.g.,][]{2006MNRAS.367...19K, Gralla2014}, and its most iconic feature is a Y-shaped current sheet separating the open and closed field line zone~\citep{Cerutti2015}. This Y-shaped current sheet can be a source of significant particle acceleration and intense high energy radiation~\citep[e.g.,][]{2014ApJ...795L..22C, Cerutti2015, Guépin2019}.

On the other hand, pulsar radio emission is believed to originate via some plasma processes from the base of the open field line zone~\citep[see e.g.][]{Lyubarsky2008}. It was recently suggested that the pair cascade process at the pulsar polar cap may directly source the observed radio emission~\citep{Philippov2020}. Results from 2D and 3D Particle-in-Cell simulations seem to support this proposition~\citep{Cruz2024, Philippov2020}. However, if this proposition is true, in order to understand the detailed properties of radio emission, it is critical to understand the microphysics of the pair-producing gap at the pulsar polar cap.

Pair production at the polar cap has been an important topic since the very beginning of pulsar research. The vacuum-gap model of \citet{Ruderman1975} proposed that the absence of charges above the polar cap leads to the formation of a strong parallel electric field, triggering bursts of electron–positron pair creation and forming discrete sparks. This scenario has been used to interpret subpulse drifting and the nested-cone beam geometry. Later extensions by \citet{GilMitra2001, GilMelikidze2003} incorporated thermionic ion flow, spark heating feedback, and soliton-induced coherent curvature radiation. Alternatively, the Space-Charge-Limited Flow model~\citep{AronsScharlemann1979} assumes continuous extraction of particles from the surface. It permits a broader range of current densities, while also leading to intermittent gaps under certain global constraints. These ideas were refined through semi-analytical and Monte Carlo methods in the 1990s and 2000s, revealing the inherently time-dependent and self-regulating nature of cascades \citep{ZhangHarding1999, Timokhin2006, Beloborodov2008, TimokhinArons2013}. Strong-field QED effects, including photon splitting and bound pair formation, were also explored by \citet{UsovMelrose1995}, showing that quantum corrections can modify both the cascade multiplicity and the observational properties, especially for magnetars.

Since the 2010s, PIC simulations have enabled global and local numerical models of pair cascades. \citet{Philippov2015, Cerutti2015, Kalapotharakos2018} demonstrated self-consistent gap formation in rotating magnetospheres, while \citet{Philippov2020} and \citet{Cruz2021} showed that non-stationary polar discharges naturally generate coherent electromagnetic waves. These waves, particularly the O-mode, were found to propagate along low-density channels and may explain observed polarization features \citep{Benacek2024a}. Recent studies also investigated the role of photon polarization in boosting pair production efficiency \citep{SongTamburini2024} and included vacuum polarization corrections to model Stokes parameters \citep{Benacek2025}.

While PIC simulations have enabled first-principles modeling of global magnetospheric dynamics and local pair cascades, they face several intrinsic limitations. In the high-multiplicity regimes typical of polar cap QED discharges, the exponential particle growth imposes strict memory and timestep constraints, limiting the attainable spatial and temporal scales. The statistical noise inherent in PIC methods further complicates the resolution of low-density regions, where electric field oscillations, wave formation, and fine-scale phase-space structures become most significant. Moreover, some key aspects of the polar cap discharge remain elusive. Although many studies have demonstrated quasi-periodic gap breakdown and identified features such as particle trapping and wave excitation, the detailed structure and long-term behavior of the discharge—including the interplay between wave–particle interaction and the formation of coherent emission—remain incompletely understood. In particular, the energy-carrying modes, timescales, and spatial profiles of returning particle fluxes have yet to be quantified from first principles. Some semi-analytic work has been done to predict the oscillation spectra and energy output~\citep{Tolman2022, Okawa2024}, but the problem is too complex to admit a fully time-dependent analytic solution.

In this work, we adopt a one-dimensional Vlasov–Maxwell kinetic model to study pair discharges over pulsar polar caps. Compared to PIC methods, which suffer from exponential particle growth and numerical noise during cascade development, the Vlasov–Maxwell approach evolves the full phase-space distribution function deterministically on a fixed grid, avoiding particle noise and offering improved accuracy and long-term numerical stability. It is particularly well-suited for capturing fine-scale wave-like features in low-density regions and for tracking periodic or metastable discharge structures. 
We developed from scratch a new Vlasov-Maxwell solver in 1D with radiation physics and pair production, and applied it to the polar cap pair cascade problem.
Our simulations reveal quasi-periodic gap formation, coherent field oscillations, and relativistic particle backflow that deposits significant energy onto the polar cap. We extract characteristic timescales, spatial structures, and electromagnetic spectra from the cascade dynamics, and estimate the surface heating rate associated with returning particles. These results offer new insights into the kinetic behavior of polar cap QED cascades, with implications for coherent radio emission mechanisms and the formation of thermal X-ray hotspots.

The remainder of this paper is organized as follows. In Section \ref{sec:method}, we present the numerical framework of our one-dimensional Vlasov–Maxwell solver, \verb|PRVMs|, including discretization methods and conservation properties. Section \ref{sec:set} details the implementation of key physical processes relevant to polar cap cascades, including curvature radiation, radiation reaction, magnetic pair creation, and boundary conditions. In Section \ref{sec:sim}, we show the main simulation results, focusing on gap dynamics, field oscillations, and particle backflow. Section \ref{sec:Discuss} provides a theoretical interpretation of these behaviors and discusses their implications for coherent radio emission and thermal X-ray hotspots.

\section{Numerical methods}\label{sec:method}
\subsection{1D Vlasov–Maxwell System}

The pair plasma at the pulsar polar cap is collisionless, and well described by the Vlasov-Maxwell system, which captures both particle dynamics and field evolution. The Vlasov equation governs the distribution function of each particle species in phase space, while the Maxwell equations govern the evolution of the electromagnetic fields.
The relativistic Vlasov equation is given by
\begin{equation}
    \partial_{t} f_s + \frac{\bm{p}}{\gamma} \cdot \nabla_{\bm{x}} f_s + q_s \left( \bm{E} + \frac{\bm{p} \times \bm{B}}{\gamma} \right) \cdot \nabla_{\bm{p}} f_s = 0,
\end{equation}
where $f_s = f_s(t, \bm{x}, \bm{p})$ is the distribution function and $s$ denotes the particle species; $t, \bm{x}=(x, y, z), \bm{p}=(p_x, p_y, p_z)$ are the time, space, and momentum respectively; $\bm{E}=(E_x, E_y, E_z), \bm{B}=(B_x, B_y, B_z)$ are the electric and magnetic fields. The Lorentz factor is $\gamma = \sqrt{1 + |\bm{p}|^2}$ in normalized units where $m_e c$ is the momentum unit. The the evolution of the electromagnetic fields is govern Maxwell equations(written in Heaviside-Lorentz units with $c=1$),
\begin{equation}
\begin{split}
    &\nabla \cdot \bm{E}=\rho,\\
    &\nabla \cdot \bm{B}=0,\\
    &\nabla \times \bm{E}=-\partial_{t}\bm{B},\\
    &\nabla \times \bm{B}=\partial_{t}\bm{E}+\bm{J}.
\end{split}  
\end{equation}

The pulsar magnetosphere can be approximated as force-free as long as there is enough charges to provide the corotating charge density $\rho_\mathrm{GJ}$. Under this approximation, the plasma is essentially massless and the net force on the plasma is zero:
\begin{equation}
    \rho \bm{E} + \bm{J} \times \bm{B} = 0, \quad \bm{E} \cdot \bm{B} = 0.
    \label{FFE}
\end{equation}

In the corotating frame of the magnetosphere satisfying FFE approximation, the Maxwell equations are expressed as
\begin{equation}
\begin{split}
    &\nabla \cdot \bm{E} = \rho -\rho_{_{GJ}}, \\
    &\partial_{t} \bm{E} = \bm{J}_0-\bm{J},
\end{split}
\end{equation}
where $\rho_{_{GJ}}=\nabla \cdot \bm{E}_{FFE}$ is Goldreich-Julian Charge Density. $\bm{J}_{0} = c\nabla\times\bm{B}$ refers to the magnetospheric current. In particular, global FFE solutions indicate that open field lines develop a toroidal component $B_\phi$ regulated by the light cylinder, which gives rise to a field-aligned current distribution,
\begin{equation}
    J_0 = c \nabla \times B_\phi,
\end{equation}
that threads the polar cap. Near the neutron star surface, the corotating and inertial frames yield nearly identical electromagnetic fields and source terms, allowing local analyses to be carried out in either frame with negligible discrepancy \citep{Gralla2020}.

In the pulsar polar cap region, owing to the intense magnetic field strength of (${\sim} 10^{12}\ G$), the momentum component of charged particles perpendicular to the magnetic field lines dissipates rapidly due to synchrotron radiation (SR). Consequently, charged particles predominantly move along the magnetic field lines. Furthermore, the radius of curvature of the magnetic field lines is approximately on the order of $10^{4}$ times the height of the gap region within the polar cap. This allows us to approximate the motion of charged particles as effectively one-dimensional. In the context of a one-dimensional system, where variation is restricted along the magnetic field line (which we take as the $x$-axis), the Vlasov–Maxwell equations simplify considerably, allowing for focused analysis of fundamental plasma interactions.

In scenarios where the background magnetic field significantly exceeds the electric field, we can assume that the magnetic field is static, meaning $\partial_{t}\bm{B}=0$; Therefore, specializing to 1D, we can reduce the set of equations to:

\begin{equation}
\begin{split}
    \partial_{t} f_s+\frac{p_x}{\gamma}\cdot &\partial_{x} f_s+ (q_s E_x)\cdot \partial_{p_x} f_s = 0,\\
    &\partial_{x} E_x = \rho-\rho_{_{GJ}}, \\
    &\partial_{t} E_x = J_0-J.
\label{revised_maxwell}
\end{split}
\end{equation}

Here the charge and current densities in the Maxwell equations can be directly calculated by integrating the distribution function over momentum space:
\begin{equation}
\begin{split}
    &\rho(t,x) = \sum_{s} q_s\int dp_x f_s,\\
    &J(t,x) = \sum_{s} q_s\int dp_x  \frac{p_x}{\gamma}f_s.
\end{split}
\end{equation}

The 1D Vlasov–Maxwell system is valuable for examining plasma dynamics where particle acceleration, wave-particle interactions, and non-linear electromagnetic effects play significant roles. By numerically solving this system, one can capture detailed structures in phase space, which are critical for understanding instabilities, wave propagation, and cascade phenomena in plasmas. Unlike particle-in-cell simulations, which rely on tracking individual particles and can suffer from statistical noise, the Vlasov–Maxwell system describes the distribution function continuously, thus avoiding noise and providing a more precise representation of fine-scale plasma dynamics.

To solve the 1D Vlasov–Maxwell system numerically, one needs to discretize the 1D1P phase space onto a finite grid consisting of cells. The discretization of space and time is carried out using the Finite Difference Time Domain (FDTD) method. The electric and magnetic fields, $\bm{E}$ and $\bm{B}$ as well as the current and charge densities, $J$ and $\rho$, are sampled on a finite grid. More specifically, the 1D1P phase space with a spatial size of $L$ and a momentum space from $-P_0$ to $P_0$ will be uniformly divided into $N_x \times N_p$ grids, then each cell would owe size of $\Delta x \ (=\frac{L}{N_x})\times \Delta p \ (=\frac{2P_0}{N_p})$. Defining $x^{i}\equiv(i-\frac{1}{2})\Delta x$ and $p^{j}\equiv j\Delta p$, the cell $C^{i,j}$ is centered at $(x^{i},p^{j})$ and has edges at $x^{i-\frac{1}{2}}/x^{i+\frac{1}{2}}$ and $p^{j-\frac{1}{2}}/p^{j+\frac{1}{2}}$. The distribution function $f_s$ was resampled in each cell:
\begin{equation}
\begin{split}
    f_s^{i,j} \equiv \frac{1}{\Delta x} \int_{x^{i-\frac{1}{2}}}^{x^{i+\frac{1}{2}}} \int_{p^{j-\frac{1}{2}}}^{p^{j+\frac{1}{2}}} dxdpf_s(x,p),
\end{split}
\label{eqn:Discretize}
\end{equation}
Characteristic plasma scales are of the order of the Debye length which depend on plasma density. The spatial grid size should be smaller than the Debye length to accurately capture the charge screening effect and resolve the microstructures.

We have developed a new GPU-based Vlasov code \verb|PRVMs| (Pulsars' Relativistic Vlasov-Maxwell Solver, pronounced as ``Plums'') from scratch to carry out this numerical procedure. The code is written in Python and open source\footnote{https://github.com/InkyDY/Prvms}. In the rest of this section and Section~\ref{sec:set}, we describe the numerical methods in detail.

Starting from the initial conditions, the equations are evolved step by step using a specified time evolution scheme. At each step, the distribution function is updated according to Equation~\eqref{revised_maxwell}, the resulting current density is calculated, and this current density is then used to update the fields themselves. We applied upwind difference method to calculate the flux ($\bm{F}$) around the cell:
 \begin{equation}
 \begin{split}
     &F_{s,x}^{n,i-\frac{1}{2},j}=\beta^{j}f_{s}^{n,i-\Theta(\beta^{j}),j},\\
     &F_{s,p}^{n,i,j-\frac{1}{2}}=q_{s} E^{i}f_{s}^{n,i,j-\Theta(q_s E^{i})}.
 \end{split}
 \end{equation}
Here the function $\Theta(x)$ is the Heaviside step function that is zero for negative values and 1 for positive values. As a result, we can write the finite difference equation as:
\begin{equation}
\begin{split}
    \frac{f_{s}^{n+1,i,j} -f_{s}^{n,i,j}}{\Delta t} &= \frac{F_{s,x}^{n,i-\frac{1}{2},j}-F_{s,x}^{n,i+\frac{1}{2},j}}{\Delta x}\\
    &+ \frac{F_{s,p}^{n,i,j-\frac{1}{2}}-F_{s,p}^{n,i,j+\frac{1}{2}}}{\Delta p}.
\end{split}
\end{equation}
The field $E^{i}$ is defined as:
\begin{equation}
    E^{i}=\frac{1}{\Delta x}\int_{x^{i-\frac{1}{2}}}^{x^{i+\frac{1}{2}}}E(x)dx.
\end{equation}
To better illustrate the boundary conditions of the extraction process (which we will discuss later), we record the information of the electric field at the boundaries of the cell so that actually $E^{i-\frac{1}{2}}$ are evolving in the simulation.
The $\bm{E}$ field evolves in each time step as
\begin{equation}
\begin{split}
    &J^{n,i-\frac{1}{2}} =\sum_{s,j} q_{s}\beta^{j}f_{s}^{n,i-\Theta(\beta^{j}),j},\\
    &{E^{n+1,i-\frac{1}{2}}=E^{n,i-\frac{1}{2}}}-\Delta t \ j^{n,i-\frac{1}{2}}.
\end{split}
\end{equation}
Since the neutron star surface is typically considered a very good conductor, we assume that the electric field at the surface is strictly zero, i.e., $E^{n,-\frac{1}{2}}=E(t,x=0)=0$.

\subsection{Validation: Two Stream Instability}
To validate the robustness of our one-dimensional Vlasov simulation code, we employed a benchmark test using the relativistic two-stream instability. This test is crucial for verifying the accuracy of kinetic plasma simulations that incorporate relativistic effects, as it challenges the code's ability to handle high-velocity particle streams and relativistic corrections within the Vlasov–Poisson framework. In our setup, the momentums of counter-streaming electron beams are $[\pm 1,0,0]^T$, and the background protons are stationary. In this situation, the dispersion relation of a relativistic two-stream instability is described as
\begin{equation}
    1 - \frac{1}{\gamma_0^3} \frac{\omega_p^2}{(\omega - kv_0)^2} - \frac{1}{\gamma_0^3} \frac{\omega_p^2}{(\omega + kv_0)^2} = 0,
    \label{ts_dispersion}
\end{equation}
where $\omega$ is the wave frequency, $k$ is the wavenumber, and $\omega_p^2=n_0e^2/m_e$ is the plasma frequency. The imaginary part of $\omega$ corresponds to the growth rate 
 of the instability. Solving Equation~\eqref{ts_dispersion} numerically, the highest growth rate corresponding to the most unstable mode are obtained as $\Gamma/\omega_p=0.297$. To calculate the most unstable mode, the length of a periodic domain $L$ is set to be $L=10c/\omega_p$, and the upper/lower limits of the momentum domain are $p/m_ec =±10$ for electrons. The number of computational cells is $1000\times 1000$. The temporal interval is given as $c\Delta t/\Delta x=1$. The implicit method is implemented with the Picard method, and the number of iterations is 20 per time step.
 
\begin{figure}[htbp]
    \centering
    \includegraphics[width=0.5\textwidth]{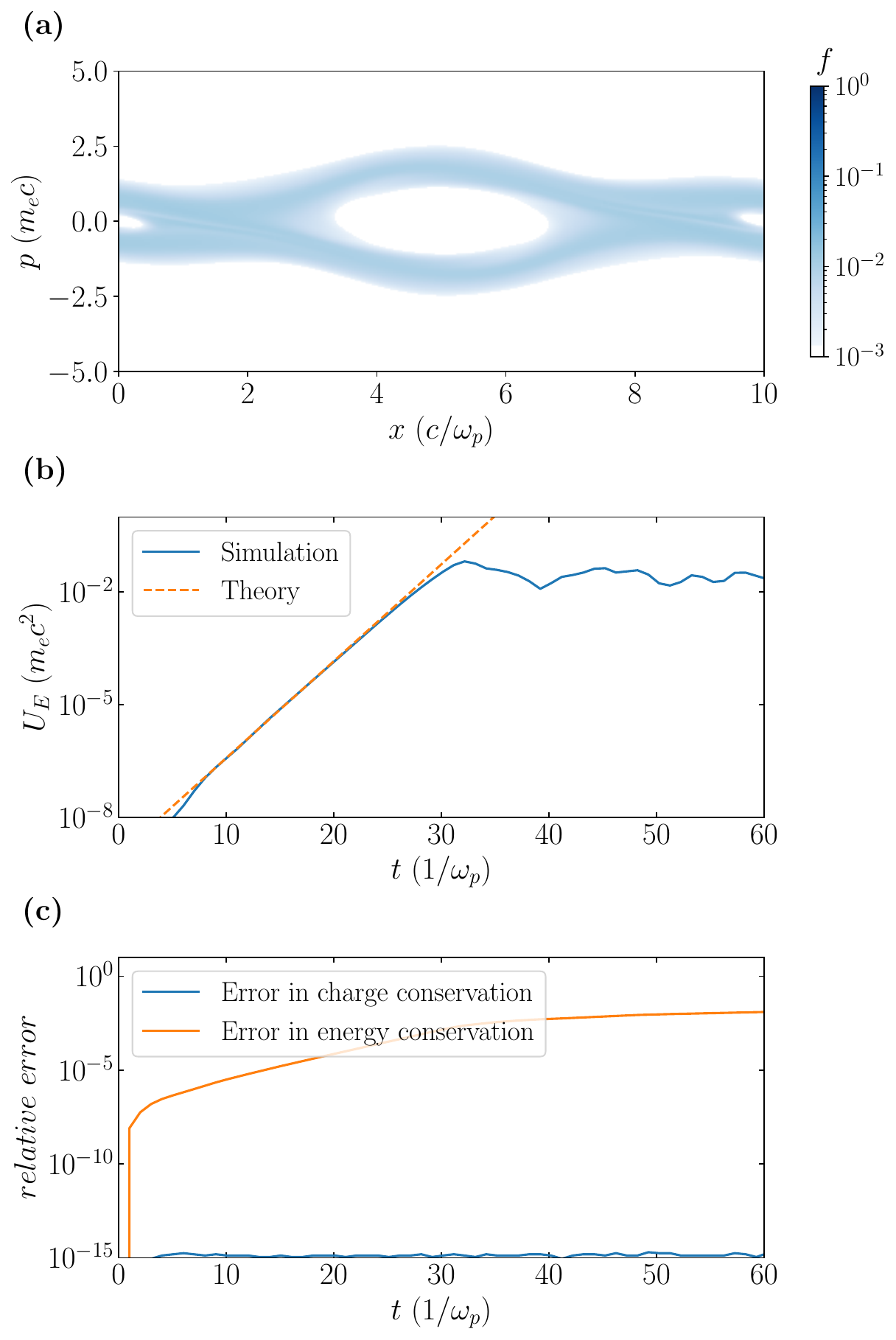}
    \caption{(a) Electron distribution function of relativistic two-stream instability solved with the proposed scheme at $t=60\ \omega_p^{-1}$. (b) The exponential amplification of the electric field energy owing to relativistic two-stream instability. The simulation reproduced the growth rate obtained from the linear theory. (c) Conservation property for relativistic two-stream instability solved with the proposed scheme. Relative error of energy conservation converges to 6\% in long-term simulation}
    \label{ts}
\end{figure}

Some previous studies \citep{Filbet2003, Shiroto2019} have already discussed schemes for the relativistic Vlasov–Maxwell system. Their proposed scheme sacrifices the conservation of the L1-norm to ensure energy conservation, e.g., the distribution function becomes negative and the conservation of L1-norm is clearly violated in the nonlinear regime. We aim to perform long-term simulations to reproduce the limit-cycle behavior of the QED cascade. Therefore, we sacrifice some accuracy to ensure that no negative values are generated, i.e., we employ the upwind difference method for the spatial and momentum differences.  Figure~\ref{ts}(a) shows the distribution function of electrons in the late nonlinear regime of the two-stream instability, indicating that the conservation of L1-norm is maintained in the discrete form and the simulation remains stable. Figure~\ref{ts}(b) illustrates the temporal evolution of the electric field energy, with time normalized by the plasma frequency ($1/\omega_p$). During the linear growth phase, the electric field energy grows exponentially, and the numerical growth rate aligns closely with the prediction from the linear theory (Equation~\eqref{ts_dispersion}). Following this phase, the amplification of the electric field energy reaches saturation, marking the transition into the nonlinear regime. Figure~\ref{ts}(c) indicates the errors of global conservation. We consider the error in energy conservation acceptable because we are conducting relativistic simulations with a large $\Delta p$, leading to an inevitable increase in relative error in momentum when approaching zero. 

The simulation results were compared with theoretical predictions derived from the relativistic linear instability analysis, and our code demonstrated a consistent agreement with the expected relativistic growth rate. This alignment verifies that our Vlasov method accurately captures relativistic effects and kinetic instabilities, confirming the code's reliability for simulating relativistic plasma dynamics. The successful replication of the relativistic two-stream instability thereby establishes the credibility of our one-dimensional Vlasov code for advanced plasma studies. 

\section{Overview of the Physical Model}\label{sec:set}
To study a one-dimensional QED cascade at a pulsar's polar cap in the context of a high-energy astrophysical environment, several additional physical processes and phenomena beyond the standard Vlasov-Maxwell system must be considered to capture the full complexity of the cascade dynamics. 

\subsection{Radiation and Radiative Reaction} 
There are three possible radiation mechanisms during the QED cascade process. Firstly, due to the curvature of magnetic field lines, electrons and positrons flowing along the background field will emit curvature radiation (CR). This is the main channel for radiative loss at the pulsar polar cap, and long believed to be the source of the $e^\pm$ pairs~\citep{Ruderman1975, DaughertyHarding1983}.
Secondly, freshly produced pairs may have finite pitch angles with respect to the background magnetic field, so they will rapidly cool via synchrotron radiation. This radiation channel only applies to secondary pairs since the primary particles are essentially trapped at the lowest Landau level due to the very short cooling time.
The last is inverse Compton (IC) scattering: relativistic particles can scatter low-energy background photons, transferring energy to the photons and contributing to the high-energy photon population. In the simulations presented in this study, we exclusively consider curvature radiation, as SR photons possess insufficient energy, and both the plasma and photon number densities are insufficient low for IC scattering to be efficient.

The charged particles with large Lorentz factors emit $\gamma$ rays of energy $\varepsilon$ at a rate of

\begin{equation}
    \frac{dN_{\gamma}}{dt d\varepsilon}=\frac{1}{\sqrt{3}\pi}\frac{e^2}{\hbar^2 c}\frac{1}{\gamma^2}\int_{\frac{\varepsilon}{\varepsilon_{c}}}^{\infty}K_{\frac{5}{3}}(x)dx,
\end{equation}
where $K_{5/3}$ is Macdonald function, also known as the modified Bessel function of the second kind, and $\varepsilon_{c}$ is the characteristic energy of the emitted curvature photon:
\begin{equation}
    \varepsilon_{c}=\frac{3}{2}\hbar \frac{c}{R_c}\gamma^3 \approx \frac{3}{2}\hbar \frac{c}{R_c} p^3.
    \label{Eq:eng_photon}
\end{equation}
To simplify the problem, it is assumed that all CR photons are emitted at the critical energy, then the total rate would be 
\begin{equation}
    \label{eqn:curvature-rate}
    \frac{dN}{dt}=\frac{4e^{2}}{9 \hbar R_c} \gamma \approx \frac{4e^{2}}{9 \hbar R_c} p.
\end{equation}

The energy loss of particles due to radiation needs to be included in the particle dynamics. After all, the momentum evolves as:
\begin{equation}
    \label{eqn:dpdt-radiation}
    \frac{dp}{dt}=q_s E - W_{rr},
\end{equation}
where $W_{rr}$ is the term responsible for radiation reaction which is opposite to the direction of $p$. For curvature radiation, it is given by
\begin{equation}
    W_{rr}\approx\frac{2e^2}{3R_{c}^{2}}p^4, 
\end{equation}
where $R_c$ is the radius of curvature of $\bm{B}$ field line at polar cap. Equation~\eqref{eqn:dpdt-radiation} needs to be taken into account in the Vlasov equation that we evolve.

To incorporate the effects of radiation reaction, we note that the evolution of the distribution function \( f \) in the momentum direction becomes significantly more complicated than in the case without radiation losses. In particular, the momentum-space dynamics is no longer trivial due to the presence of the radiation damping term, i.e. the advection velocity of \( f \) in the momentum direction, given by \( eE^{\text{i}} -2e^2/3R_{c}^{2} \cdot {p^{j}}^{5}/|p^{j}| \), is no longer uniform, so we can no longer approximate the flux at cell boundaries by using the cell-averaged \( f \) multiplied by a constant velocity, as this simplification neglects the momentum dependence of the advection velocity and may result in significant cumulative discretization errors, especially in regions where \( f \) exhibits strong gradients.

To manage this complexity, we first apply the Strang-split technique~\citep[see e.g.][]{qin2002} to separate the evolution of \( f \) along spatial and momentum directions. This operator-splitting step simplifies the numerical treatment, allowing us to focus on the more challenging momentum-direction dynamics in isolation. Since the spatial advection is trivial in our case, we focus our attention on the more involved momentum-direction evolution. Consider the effect of the momentum advection,
\begin{equation}
\begin{split}
    \frac{df}{dt}&=\frac{\partial f}{\partial t}+\frac{d p}{d t} \frac{\partial f}{\partial p}=0,
\end{split}
\end{equation}
where $dp/dt$ is given by Equation~\eqref{eqn:dpdt-radiation}. By using the method of characteristics, the characteristic curves are the solutions of the ODE:
\begin{equation}
    \frac{d p^{j}}{d t}={eE^{i}}-\frac{2e^2}{3R_{c}^{2}} \frac{p^{j^{5}}}{|p^{j}|}.
\label{eqn:characteristiccurves}
\end{equation}
We integrate the characteristic curves using a fourth-order Runge-Kutta (RK4) method.

The schematic of the update process is:
\begin{equation}
f^{n,i,j} \xrightarrow{\hat{P}(E^i)} f^{n+1,i,j'} \rightarrow f^{n+1,i,j}, 
\end{equation}
Here, $\hat{P}(E^i)$ denotes the projection of the distribution function along characteristics determined by the electric field $E^i$, and $f^{n+1,i,j'}$ represents the intermediate distribution function at time \(n+1\) on a distorted momentum grid resulting from the characteristic flow.

To compute $f^{n+1,i,j'}$, we conserve the phase-space density along the characteristic curves. The plasma distribution function $f^{n,i,j}$ within the cell $C^{i,j}$ is transported along the trajectory given by Equation~(\ref{eqn:characteristiccurves}).  For each momentum grid point, the characteristic equation is solved over a time step $\Delta t$ to trace the departure point in momentum space. Under the characteristic flow, the original momentum cell boundaries $p^{j \pm 1/2}$ are mapped to $p^{j' \pm 1/2}$, forming a deformed cell in which the distribution is updated. The updated distribution function in the deformed cell is then given by
\begin{equation}
f^{n+1,i,j'} = \frac{p^{j + \frac{1}{2}} - p^{j - \frac{1}{2}}}{p^{j' + \frac{1}{2}} - p^{j' - \frac{1}{2}}} f^{n,i,j}, 
\end{equation}
where the prefactor accounts for compression or dilation of the momentum cell under the characteristic transformation. Finally, the function is projected back onto the original grid using Equation~(\ref{eqn:Discretize}) to obtain $f^{n+1,i,j}$.

\subsection{Magnetic Pair Production}
Magnetic pair production ($\gamma\xrightarrow{\textbf{B}}e^-+e^+$ ) is a quantum electrodynamic (QED) process that occurs in the presence of ultra-strong magnetic fields, such as those found near pulsars or magnetars~\citep{Erber1966, DaughertyHarding1983, BaringHarding2001}. This phenomenon involves the conversion of a high-energy photon into an $e^{\pm}$ pair when the photon interacts with the intense magnetic field. The cross-section of the photon absorption is given by~\cite{Erber1966}:
\begin{equation}
    \frac{d\sigma}{dz} \approx 0.23 \frac{B}{B_Q} \sin\theta \, \alpha \, \lambda_c \exp\left(-\frac{8}{3\chi}\right) \Theta(\varepsilon_{\gamma} \sin\theta - 2),
\end{equation}
where $\chi \approx (B/B_Q) \varepsilon_{\gamma} \sin\theta$, $B_Q = \frac{m_e^2 c^3}{e \hbar} \approx 4.41 \times 10^{13} \mathrm{G}$ {is the Schwinger magnetic field}, and $\alpha$, $\lambda_c$ are the fine-structure constant and the Compton wavelength of the electron, respectively. $\theta$ is the angle between the photon propagation direction and the magnetic field. The function $\Theta(x)$ is the Heaviside step function that is zero for negative values and 1 for positive values. In our simulations, CR photons are emitted tangentially to the magnetic field lines, and the angle increases linearly with distance from the emission point, even though we follow their motion as if they travel along the field lines as well. The Heaviside step function ensures the threshold condition for pair production, $\varepsilon_{\gamma} \sin \theta \geq 2$. 

The Monte Carlo technique can be utilized to model the stochasticity of the radiation transfer problems including CR and pair production. However, we adopt a simplified version to reduce computational costs while maintaining a high-resolution simulation: we assume the CR photons are only produced at the critical energy, and then the emission rate is proportional to the momentum of the parental charged particle 
as given by Equation~\eqref{eqn:curvature-rate}.
For pair production, we assume it would happen once $\varepsilon_{\gamma}$ sin$\theta> 2$ is satisfied, so the effective free path is $\ell_{\rm ph} \approx 2 R_c/\varepsilon_\gamma$ since $R_c \gg \ell_\mathrm{ph}$. We also assume that the generated $e^\pm$ pairs evenly share the energy of parent photons. In other words, the transfer rate equations are simplified to
\begin{equation}
    \Gamma_{1}(p,p',\theta)=\delta\left(p-\frac{1}{2}p'\right)\delta\left(\theta-\frac{2}{p'}\right),
\end{equation}
\begin{equation}
    \Gamma_{2}(p,p')=\frac{4}{9} \frac{e^2/R_c}{\hbar} p' \delta\left(p-\frac{3}{2}\frac{\hbar}{R_c} p'^3\right).
\end{equation}
Overall, the 1D Vlasov-Maxwell system that we are solving can be expressed as 
\begin{equation}
\label{eqn:vlasov-full}
\begin{split}
    \partial_{t} f_s+\frac{p_x}{\gamma}\cdot \partial_{x} f_s+& (q_s E_x-W_{rr})\cdot \partial_{p_x} f_s = Q_\mathrm{pp}+Q_\mathrm{ext},\\
    &\partial_{x} E = \rho -\rho_{_{GJ}}, \\
    &\partial_{t} E = J_0-j.
\end{split}
\end{equation}
There are two source terms representing the injection of particles into the system. $Q_\mathrm{ext}$ represents extraction of charges from the stellar surface, while $Q_\mathrm{pp}$ represents pair production:
\begin{equation}
    Q_\mathrm{pp}(t,x,p)=\iint dp'\ d\theta \Gamma_{1}(p,p',\theta)f_{\gamma}(t,x,p',\theta).
\end{equation}
Here, we must record the distribution function $f_{\gamma}$ of photons within the system which is in a 1D1V1A phase space (A for the angle $\theta$ with respect to the local magnetic field). It evolves with time under its own Vlasov equation:
\begin{equation}
    \partial_{t} f_{\gamma} + c\ \partial_{x}f_{\gamma} + \frac{c}{R_c}\partial_{\theta}f_{\gamma} = Q_\mathrm{rad}(t,x,p,0),
\end{equation}
where
\begin{equation}
\begin{split}
    Q_\mathrm{rad}(t,x,p,0)=&\sum_{s}\left[\int_{P_{th}}^{P_{0}}dp' \Gamma_{2}(p,p') f_{s}(t,x,p')\right.\\
    &\left.+\int_{-P_{0}}^{-P_{th}}dp' \Gamma_{2}(p,p') f_{s}(t,x,p') \right].
\end{split}
\end{equation}
Notably, if we record the position, momentum, and angle of photons, the computational complexity of \( f_{\gamma} \) will be \( O(N_p N_x N_A) \sim O(N_x^3) \), since momentum, which corresponds to energy, determines the free path. Furthermore, since the angle variable \( A \) is directly proportional to the photon travel time \( t_{\text{travel}} \), we reparameterize the angle dependence in terms of the remaining propagation time \( t - t_{\text{current}} \). This transformation enables us to replace the angular dimension with a time-based lookup. Consequently, we precompute a list of pair injection events indexed by time and phase space, specifying when and where pairs are to be deposited in the simulation domain. This approach eliminates the need for explicit iteration over all angles at each timestep, while still faithfully preserving the angular dependence of the injection process. If we further consider that the momentum of CR photons is significantly lower than that of the parental charged particles (typically less than \( 100~d_p \)), the complexity reduces to roughly \(\mathcal{O}(N_x^2)\), inline with the non-radiative Vlasov update.

\section{1D simulations of QED cascade}\label{sec:sim}

We apply the Vlasov code described in Sections~\ref{sec:method} and~\ref{sec:set} to study the pulsar polar cap cascade problem in two opposite regimes:
the Ruderman–Sutherland  model and the Space-Charge-Limited Flow model.

The RS model is applicable in situations where the Goldreich–Julian charge density $\rho_{\mathrm{GJ}} > 0$, such that ion extraction from the surface is required to supply the magnetospheric current \citep{Ruderman1975}. 
If the work function and surface gravity is strong enough, ions cannot be readily extracted. As a result, a vacuum gap naturally forms above the surface, within which strong electric fields build up and initiate pair creation. In our simulation, we mimic this boundary condition by disallowing the extraction of charged particles from the stellar surface, effectively modeling the scenario in which only pair creation (rather than surface emission) populates the magnetosphere \citep{Timokhin2010}. 
Practically, we set $Q_\mathrm{ext} = 0$ in Equation~\eqref{eqn:vlasov-full}, and all plasma in the open field line region must be produced locally through QED pair cascades described by the $Q_\mathrm{pp}$ term.
Once a gap with an unscreened parallel electric field forms near the surface, returning positrons (or electrons) from previous cycles are re-accelerated, emit curvature radiation, and trigger fresh pair production. The resulting plasma then screens the electric field, halting further acceleration, and flows outward. A new discharge is initiated only after the escaping plasma density drops sufficiently so that $n_\pm \lesssim n_\mathrm{GJ}$, at which point the field can re-emerge. This quasi-periodic behavior forms a limit cycle with sharply defined gap formation, rapid pair bursts, and intervening quiet phases. Despite its idealized boundary condition, the RS model has proven indispensable for understanding the basic physics of strong QED cascades in time-dependent regimes.

The behavior of the space-charge-limited flow model, on the other hand, depends critically on the dimensionless parameter \(\eta = J_0 / J_{\rm GJ}\), where \(J_0\) is the current density imposed by the global magnetosphere and \(J_{\rm GJ}\) is the local Goldreich–Julian current density. When \(0 < \eta < 1\), a steady-state analytic solution exist and describe an oscillatory, mildly relativistic flow.  However, the numerical investigations \citep{Chen2012, TimokhinArons2013} show that such solutions are dynamically unstable. The flow quickly evolves into a time-dependent configuration consisting of a warm, low-energy beam and a cloud of trapped charges of the same sign. These trapped particles adjust the local charge density to match \(\rho_{\rm GJ}\), while the beam maintains the imposed current. Since the maximum particle energies are insufficient to initiate pair cascades, therefore unable to sustain radio emission, this regime results in a so-called ``dead zone'' in the pulsar polar cap. In contrast, when \(\eta > 1\), no steady-state solution exists. The mismatch between current and available charge density leads to strong acceleration and the development of a high-voltage gap beginning at the stellar surface. For \(\eta \gg 1\), the gap undergoes quasi-periodic breakdowns through bursts of pair creation, producing a limit-cycle discharge behavior.

Direct 1D PIC simulations have demonstrated that both RS and SCLF regimes can exhibit strongly time-dependent discharge behavior when the magnetospheric current density differs appreciably from \(J_\mathrm{GJ}\)~\citep{Timokhin2010, TimokhinArons2013}. In this paper, we follow the previous 1D PIC simulations and setup the RS and SCLF regimes as follows. The RS model corresponds to the limiting case where no surface particles are supplied, while the SCLF model allows a tunable degree of surface particle extraction. In our simulation framework, this tunability is introduced via a parameter \(\zeta\), which determines the fraction of induced surface charge injected as cold particles. In our numerical setup, we assume that the neutron star surface is located at a position within the interval \( [0, \Delta x] \). The boundary condition is set such that the electric field below the surface, \( E_{x=0} \), is identically zero. This implies that the induced surface charge density within the first cell must satisfy the relation
\begin{equation}
    \rho = \rho_{\text{GJ}} - \rho^{0} + \frac{E_{\Delta x} - E_0}{\Delta x},
\end{equation}
where \( \rho^0 \) represents the charge density that originally existed within the first spatial interval \([0, \Delta x]\), before the adjustment from the boundary condition. Then a fraction \( \zeta \) of the induced charge \( \rho \) is assigned to cold particles with zero initial momentum, placed at \( x = 0 \). The system's self-consistent electric field then determines which particles are accelerated into the simulation domain and which remain trapped near the boundary due to insufficient field strength. This approach permits continuous interpolation between RS-like and SCLF-like boundary conditions.

We adopt canonical pulsar parameters to initialize the system: magnetic field strength \( B \sim 10^{12}\,\mathrm{G} \), stellar radius \( R_\star \sim 10^6\,\mathrm{cm} \), and the spin period \(P \sim 2s\) so that light-cylinder radius \( R_\mathrm{LC} \sim 10^{10}\,\mathrm{cm} \). The curvature radius of magnetic field lines is estimated as \( R_c \sim \sqrt{R_\star R_\mathrm{LC}} \sim 10^8\,\mathrm{cm} \), and the Goldreich–Julian number density is \( n_\mathrm{GJ} \sim 10^{12}\,\mathrm{cm}^{-3} \). The polar cap radius $r_\mathrm{pc}$ is given by $r_\mathrm{pc} \sim R_*\sqrt{R_*/R_\mathrm{LC}} \sim 10^4\,\mathrm{cm}$.

\begin{table}[ht]
\centering
\caption{Key simulation parameters and rescaled units used in simulations.}
\begin{tabular}{ll}
\toprule
\textbf{Parameter} & \textbf{Value / Description} \\
\midrule
Magnetic field strength \( B \) & \( 10^{12}\,\mathrm{G} \) \\
Neutron star radius \( R_\star \) & \( 10^6\,\mathrm{cm} \) \\
Light cylinder radius \( R_\mathrm{LC} \) & \( 10^{10}\,\mathrm{cm} \) \\
Curvature radius \( R_c \) & \( \sqrt{R_\star R_\mathrm{LC}} \sim 10^8\,\mathrm{cm} \) \\
Unit number density \( n_0 \) & \( \sim 10^{12}\,\mathrm{cm}^{-3} \) \\
Plasma frequency \( \omega_p \) & \( \sim 10^{10}\,\mathrm{s}^{-1} \) \\
Unit time \( t_0 = \Delta t \) & \( 1/\omega_p \sim 10^{-10}\,\mathrm{s} \) \\
Unit length \( x_0 = \Delta x \) & \( c t_0 \sim 1\,\mathrm{cm} \) \\
Unit electric field \( E_0 \) & \( m_e c \omega_p / e \) \\
Polar cap radius \( r_{pc} \) & \(  10^4\,\mathrm{cm} \) \\
\midrule
\multicolumn{2}{l}{\textbf{Rescaled quantities}} \\
\cmidrule(lr){1-2}
Rescaled \( R_c' \) & \( 10^{-3} R_c \) \\
Rescaled \( \hbar' \) & \( 10^{-2} \hbar \) \\
Rescaled \( e' \) & \( 10^2 e \) \\
Reduced \( p_\text{th} \) & \( \sim 10^5 \) \\
\bottomrule
\end{tabular}
\label{tab:rs_parameters}
\end{table}

To reduce numerical complexity, we normalize physical quantities using characteristic plasma units:
\begin{itemize}
    \item Time unit: \( t_0 = \omega_p^{-1} \sim 10^{-10}\,\mathrm{s} \), where \( \omega_p = \sqrt{4\pi n_0 e^2 / m_e} \sim 10^{10}\,\mathrm{s}^{-1} \),
    \item Length unit: \( x_0 = c t_0 \sim 1\,\mathrm{cm} \),
    \item Electric field unit: \( E_0 = m_e c \omega_p / e \).
\end{itemize}
In the rest of this paper, we will use the subscript $0$ (e.g.\ ${x}_0$) to denote the unit of the corresponding quantity, and a tilde (e.g.\ $\tilde{x}$) to denote dimensionless quantities, $\tilde{x} \equiv x/x_0$.

While these natural units preserve physical fidelity, they lead to extremely low pair production rates:
\begin{equation}
    \frac{dn}{dt} \sim \frac{4 e^2}{9 \hbar R_c} p \sim 10^{-10} p \cdot \frac{dn_0}{dt_0},
\end{equation}
which imposes prohibitively long simulation runtimes. To overcome this, we introduce controlled rescaling of key physical constants (see Table~\ref{tab:rs_parameters}):
\begin{equation}
    R_c' = 10^{-3} R_c, \quad \hbar' = 10^{-2} \hbar, \quad e' = 10^2 e.
\end{equation}
These scalings enhance the pair production rate by five orders of magnitude and reduce the threshold momentum for radiation-reaction balance from \( p_\mathrm{thr} \sim 10^8 \) to \( p_\mathrm{thr} \sim 10^5 \), making the momentum space more tractable numerically.

The simulation domain length \( L \sim 10^3\,\mathrm{cm} \) is chosen to match the physical height \(H\) of the QED cascade region above the pulsar surface. This scale is not merely a numerical setting but carries direct physical significance: it determines whether a full discharge cycle can unfold within the cascade zone. A successful discharge requires a complete sequence of electric field growth, particle acceleration, pair production, and screening to take place before particles escape. Since the electric field can only regenerate after the plasma exits the domain, \( H \) sets the cadence of the entire limit-cycle evolution. Note that \( H\) needs to be smaller than the polar cap radius, $H \lesssim r_\mathrm{pc}$, for the 1D approximation to hold, otherwise geometric effects will start to affect the profile of $E_\parallel$ in the gap~\citep{TimokhinArons2013}.

\begin{figure*}[htbp]
    \centering
    \includegraphics[width=\textwidth]{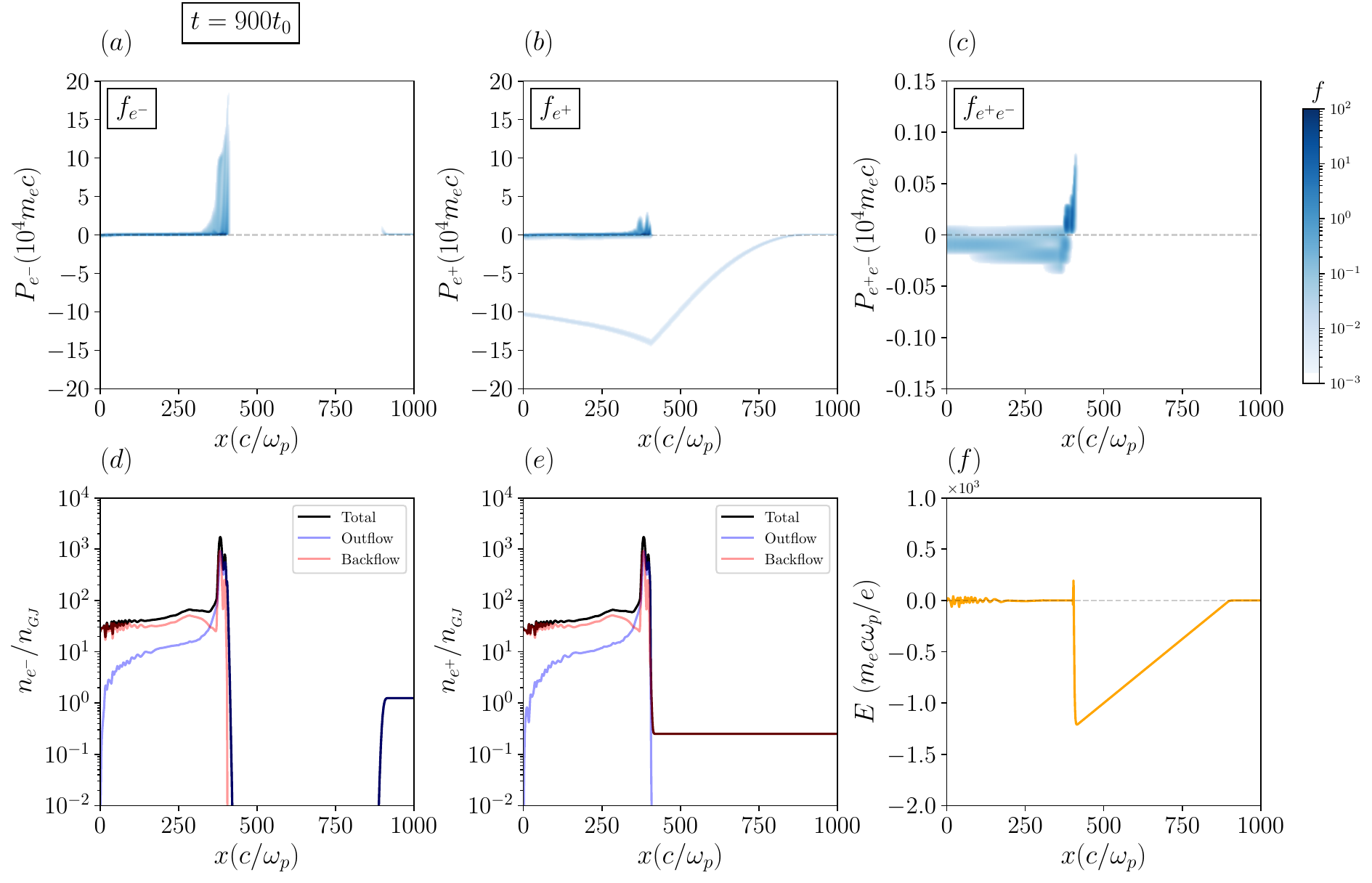}
    \caption{Snapshot of the simulation of the RS case at \( t = 900 t_0 \) (\(n_{\text{GJ}}=2n_0, \eta=1.5\)).
    Panels (a)–(c): Phase space distributions of electrons, positrons, and newly produced pairs.
    Panels (d)–(e): Normalized number densities.
    Panel (f): Electric field profile.\\
    The same panel layout is also used for the simulations of the SCLF model.\\ 
    A full animation of the temporal evolution can be viewed at: \href{https://youtu.be/3NRAFCCmY4w}{RS-movie}}
    \label{fig:RS_snapshot}
\end{figure*}

We focus on how the following four physical quantities influence the structure and dynamics of the QED discharge:
\begin{enumerate}
    \item Curvature radius \( R_c \): controls the efficiency of curvature radiation and the energy of emitted photons;
    \item Magnetic field strength \( B \): determines whether curvature photons can exceed the threshold for magnetic pair production;
    \item Magnetospheric current density \( J_0 \): expressed via the dimensionless ratio \( \eta \equiv J_0 / J_\mathrm{GJ} \), it regulates whether the system remains steady or becomes time-dependent;
    \item Cascade region height \( H \): sets the physical length over which the gap can evolve and discharge cycles can fully develop.
\end{enumerate}

These four parameters govern the feedback loop between electric field growth, particle acceleration, radiation emission, pair injection, and field screening. They also shape the formation of plasma blobs and backflow, which are key to surface heating and observable polar cap activity. Table~\ref{tab:rs_parameters} summarizes key simulation parameters and rescaled quantities. In the rest of this section, we will present numerical simulations of both the RS model and the SCLF model. At the same time, we try to develop a quantitative model to describe the behavior of the QED cascade, using the simulations as guidance.

\begin{figure*}[htbp]
    \centering
    \includegraphics[width=\textwidth]{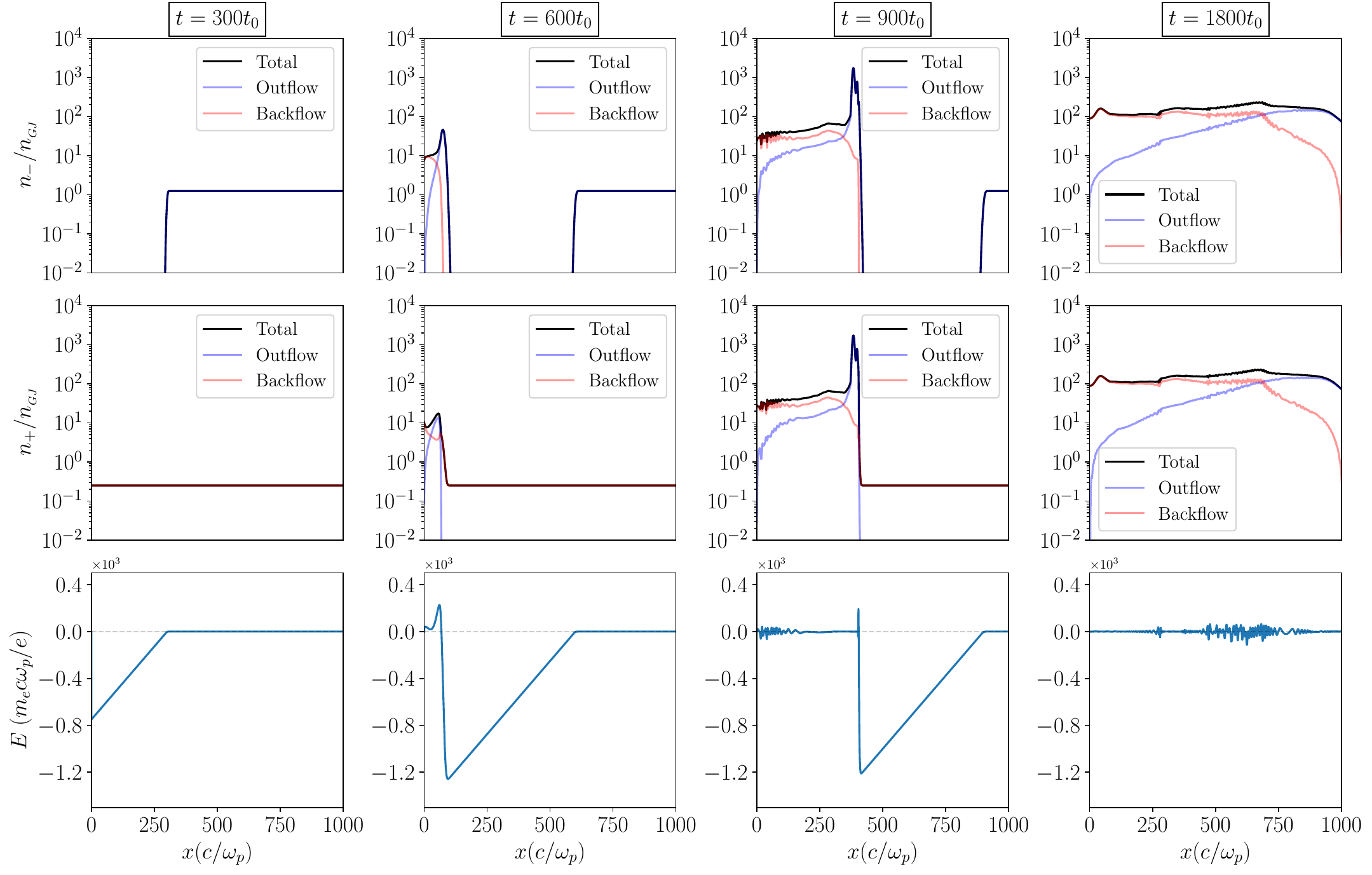}
    \caption{
    Temporal evolution of the Ruderman–Sutherland discharge at four selected snapshots: 
    \( t = 300\, t_0 \), \(600\, t_0\), \(900\, t_0\), and \(1800\, t_0 \) (left to right). 
    Top/Middle row: normalized electron/positron number density.
    Bottom row: electric field profile. 
    Outflow and backflow components are shown separately in each density panel.
    These panels illustrate the cyclic dynamics of the RS model. 
    The first column shows the initial development of the electric field in the absence of charges. 
    The second column captures the formation of a vacuum gap. 
    The third column shows the propagation of the gap and the onset of a QED cascade near its left edge. 
    The fourth column depicts the relaxation phase, in which the electric field is screened and the system prepares for the next discharge cycle.
    }
    \label{fig:RS_stages}
\end{figure*}

\subsection{Simulations of the RS Model}

We begin by examining the overall structure of plasma evolution in the Ruderman–Sutherland regime. In this case, the neutron star surface does not supply any plasma; all charges must be generated self-consistently through the QED discharge process. The magnetospheric current density is chosen to be slightly super-Goldreich–Julian, with \( \eta = J_0 / J_\mathrm{GJ} = 1.1 \). This is motivated by the fact that $\eta > 1$ is expected in the central regions of the polar cap due to general relativistic effects~\citep{Muslimov1992}.

The simulation is initialized with a tenuous pair plasma that satisfies both the local charge density \( \rho = \rho_\mathrm{GJ} \) and the total current density \( j = J_0 \). The number densities of electrons and positrons are given by:
\begin{equation}
n_e = \frac{\eta + 1}{2} n_\mathrm{GJ}, \quad
n_p = \frac{\eta - 1}{2} n_\mathrm{GJ}.
\end{equation}

These are the minimum number densities required to satisfy the conditions. These particles are initialized as counter-streaming cold beams with momenta \( \pm p_0 \), where \( p_0 = 50\,m_e c \). This ensures that the velocities of both species are close to $c$ which allows the system to satisfy $j = J_0$. The electrons propagate outward, while the positrons travel inward toward the stellar surface.

This configuration initially carries the required magnetospheric current while satisfying $\rho = \rho_\mathrm{GJ}$. However, because the Ruderman–Sutherland boundary prohibits any surface particle supply, the initial plasma soon separates as the particles propagate. As a result, the charge and current balance near the surface can no longer be maintained, forming a gap with an unscreened electric field. The system then enters a sequence of stages characterized by field growth, particle acceleration, pair creation, and plasma screening, which repeats in a self-regulated limit cycle, see Figure~\ref{fig:RS_stages}.

\textbf{Stage I - Formation of the Gap}:

As the initial electron beams propagate, the surface region becomes depleted of plasma. Since no particles can be extracted from the surface under the RS boundary condition, an unscreened electric field starts to develop. The time-dependent electric field profile on the left side (near $x = 0$) evolves approximately as:
\begin{equation}
    E(x, t) = \frac{1 + \eta}{2\eta} J_0(\beta t - x) \, \Theta(\beta t - x),
\end{equation}
where $\beta \sim c$ is the front speed of the growing gap. This behavior is consistent with the early-time gap development described by~\citet{2015ApJ...810..144T}, who found similar electric field profiles under the RS boundary condition. In the following, we largely adopt the same theoretical
approach.

To estimate the characteristic gap length, consider the positron beams that fall back toward the surface under the influence of the electric field. Their momentum profile is given by:
\begin{equation}
    p=\int E dt =\frac{1}{4}\frac{1 + \eta}{2\eta} J_0(t-x/\beta)^2\, \Theta(\beta t - x).
\end{equation}

These initial positrons, referred to as Gen~0, are accelerated by the unscreened electric field and emit curvature radiation photons. Among them, the earliest and most efficiently accelerated particles produce the first photons energetic enough to trigger pair creation. The position at which the first such Gen~0 particle generates a pair defines the effective size of the gap.

The total gap size, \( l_{\text{gap}} \), can be decomposed into two components: the acceleration length of the returning positrons, \( l_p \), and the mean free path of the curvature radiation photons, \( l_{\gamma} \). As established above, the positron momentum \( p_+ \) scales as \( p_+ \propto l_p^2 \), and the energy of a single curvature photon scales as \( \varepsilon_{\gamma} \propto p_+^3 \). Since the photon mean free path is inversely proportional to its energy, it follows that
\begin{equation}
    l_{\gamma} \propto \varepsilon_{\gamma}^{-1} \propto l_p^{-6}.
\end{equation}
Minimizing the total gap length with respect to \( l_p \) (i.e., setting \( \delta l_{\text{gap}} = 0 \)), we find the optimal relation
\begin{equation}
    l_p = 6 l_{\gamma}.
\end{equation}
Substituting this into the expression for the total gap size yields:
\begin{equation}
    l_{\text{gap}} = 2(l_p + l_{\gamma}) = \frac{7}{3} l_p.
\end{equation}
The prefactor of 2 accounts for the continuous expansion of the gap front. It should be noted that this estimate neglects the timescale over which the peak electric field on the left side of the gap becomes screened, and thus slightly underestimates the actual value of \( l_{\text{gap}} \). By incorporating Equation (\ref{Eq:eng_photon}) and 
\begin{equation}
    l_{\gamma}\simeq\chi R_c
\frac{B_Q}{B}\frac{m_ec^2}{\varepsilon_c},
\end{equation}

we obtain the following scaling for the gap length:
\begin{equation}
\begin{split}
    l_{\text{gap}}^{\text{RS}} &\simeq 2\times10^{4} R_{c,7}^{2/7} B_{12}^{-1/7} \chi^{1/7} n_{\text{GJ}}^{-3/7} (\eta + 1)^{-3/7}\ \text{cm}\\
    &\simeq 10^{4} B_{12}^{-4/7} \chi^{1/7} P_1^{4/7} (\eta + 1)^{-3/7}\ \text{cm}.
\end{split}
\label{lgap}
\end{equation}
where \(R_{c,7}\equiv R_c/10^7 cm\), $P_1\equiv P/1\,\mathrm{s}$, and \( B_{12}\equiv B/10^{12} G\).
The characteristic timescale associated with this stage is approximately given by \( l_{\text{gap}} / c \). At $\eta > 1$, this estimate is somewhat smaller than the typical polar cap size due to the $(\eta + 1)^{-3/7}$ factor, and our 1D approximation is valid.

\textbf{Stage II - Propagation of the Gap}:

Once the gap is formed, it does not remain stationary at the pulsar surface. As Gen~1 particles begin to screen the electric field on the left side, Gen~1 electrons are also accelerated to the right to nearly the speed of light. Because the screening occurs over a finite timescale, these electrons remain exposed to the unscreened electric field long enough to gain significant energy. They continue contributing to the screening process while emitting curvature radiation photons that generate new electron-positron pairs.

The newly produced electrons are co-moving with the left boundary of the gap and are accelerated beyond the threshold momentum, repeating the same pair production mechanism. This initiates a sustained QED cascade that moves with the gap. As a result, the gap front propagates rightward at approximately the speed of light, leaving behind a high-density plasma trail.

At this stage, two physical effects become apparent. First, the electric field on the left side of the gap exhibits oscillatory behavior due to overscreening. These waves appear to propagate faster than the speed of light. In 1D geometry, this superluminal phase velocity simply reflects the rapid activation of electrons across the screening region, without violating causality. Such behavior is consistent with the spontaneous excitation of superluminal O-mode electromagnetic waves during pair discharges, as reported in \cite{Philippov2020}. In Section~\ref{Spectrum}, we perform a Fourier analysis of the entire cascade process and confirm the presence of superluminal wave structures. However, as noted by \citet{Rafat2019b}, Langmuir-like waves can appear superluminal when Lorentz-boosted from the plasma rest frame. We defer a more detailed discussion about wave modes to Section~\ref{Spectrum}.

Second, the plasma clump just left of the gap shows an initial exponential growth in density, followed by a linear phase. During the exponential stage, the electric field is not yet fully screened, allowing newly created pairs to be rapidly accelerated to energies sufficient for further pair production. The pair production rate in this regime is approximately proportional to the product \(n_e \left< \gamma\right>\) and the particle number grows exponentially. This behavior is qualitatively consistent with the analytical model proposed by~\cite{Okawa2024}.  This efficient feedback drives rapid plasma buildup. As the local density approaches \( \sim 10\, n_{\mathrm{GJ}} \), the field becomes strongly screened and eventually overscreened. Some electrons are then reflected or escape leftward. Once the electric field is fully suppressed, the newly produced electrons can no longer be accelerated efficiently. Nevertheless, the previously accelerated high-energy particles continue to emit curvature photons and generate pairs, resulting in a slower, nearly linear increase in plasma density. This phase persists until the gap reaches the outer magnetosphere, leaving behind a dense plasma population that lacks sufficient energy to sustain further pair production via curvature radiation. The characteristic timescale for this stage is approximately \( H / c \), where \( H \) is the height of the cascade region, corresponding to the box size in our simulations.

\textbf{Stage III - Relaxation of the System}:

After the gap advects out of the box, the simulation domain is filled with high-density plasma generated by the preceding QED cascade. This plasma either flows back toward the pulsar surface or escapes outward into the magnetosphere. In response to these bulk flows, the electric field exhibits residual oscillations, but no further pair production occurs.

Without new sources of plasma, the particle density gradually decreases. This depletion proceeds as plasma escapes from both ends of the domain at a characteristic rate of \( \bar{n}c \). The decay continues until the average density returns to a level comparable to the background value \( n_{\mathrm{GJ}} \), at which point the conditions are again favorable for gap formation and the onset of a new discharge cycle.

The characteristic decay timescale in this relaxation phase can be approximated as:
\begin{equation}
    t_{\mathrm{decay}} \sim \frac{H}{2c} \ln \left( \frac{\mathcal{N}}{n_{\mathrm{GJ}}} \right),
\end{equation}
where \( \mathcal{N} \) is the average plasma density left behind after the peak-density plasma blob exits the simulation domain. This quantity characterizes the residual background charge density established by each discharge and serves as a more physically meaningful measure of the long-term plasma loading. After time $t_\mathrm{decay}$, the plasma density becomes comparable to $n_\mathrm{GJ}$, and charges start to separate again to form the gap.

The total period of the limit-cycle behavior, including gap formation, cascade propagation, and system relaxation, is the sum of the timescales associated to the three stages:
\begin{equation}
    T_{RS} \approx \frac{1}{c} \left[ l_{\text{gap}}^{\text{RS}} + H \left( 1 + \frac{1}{2} \ln \left( \frac{\mathcal{N}_{\text{RS}}}{n_{\mathrm{GJ}}} \right) \right) \right].
\end{equation}

This cycle repeats as the system re-enters the charge-starved state required to initiate the next gap.

\subsection{Simulations of the SCLF Model}
We now switch our attention to the Space-Charge-Limited Flow model, where the charged particles, primarily electrons, are freely emitted from the neutron star's surface without any hindrance, allowing the electric field at the surface to be effectively screened.  In our simulations, the mechanism for charge release at the neutron star surface is not implemented in an entirely ``free emission'' fashion. Specifically:
\begin{enumerate}
    \item We limit the surface to emit a maximum cold electron flux of \( n_{\text{GJ}} c \), where \( n_{\text{GJ}} \) is the Goldreich--Julian number density.
    \item We allow for a tunable fraction \(\zeta\) of the induced surface charge to be released from the stellar interior. This is intended to model the possibility that electrons may be partially bound by physical effects such as strong nucleon-electron coupling near the neutron star surface.
\end{enumerate}

\begin{figure*}[htbp]
    \centering
    \includegraphics[width=\textwidth]{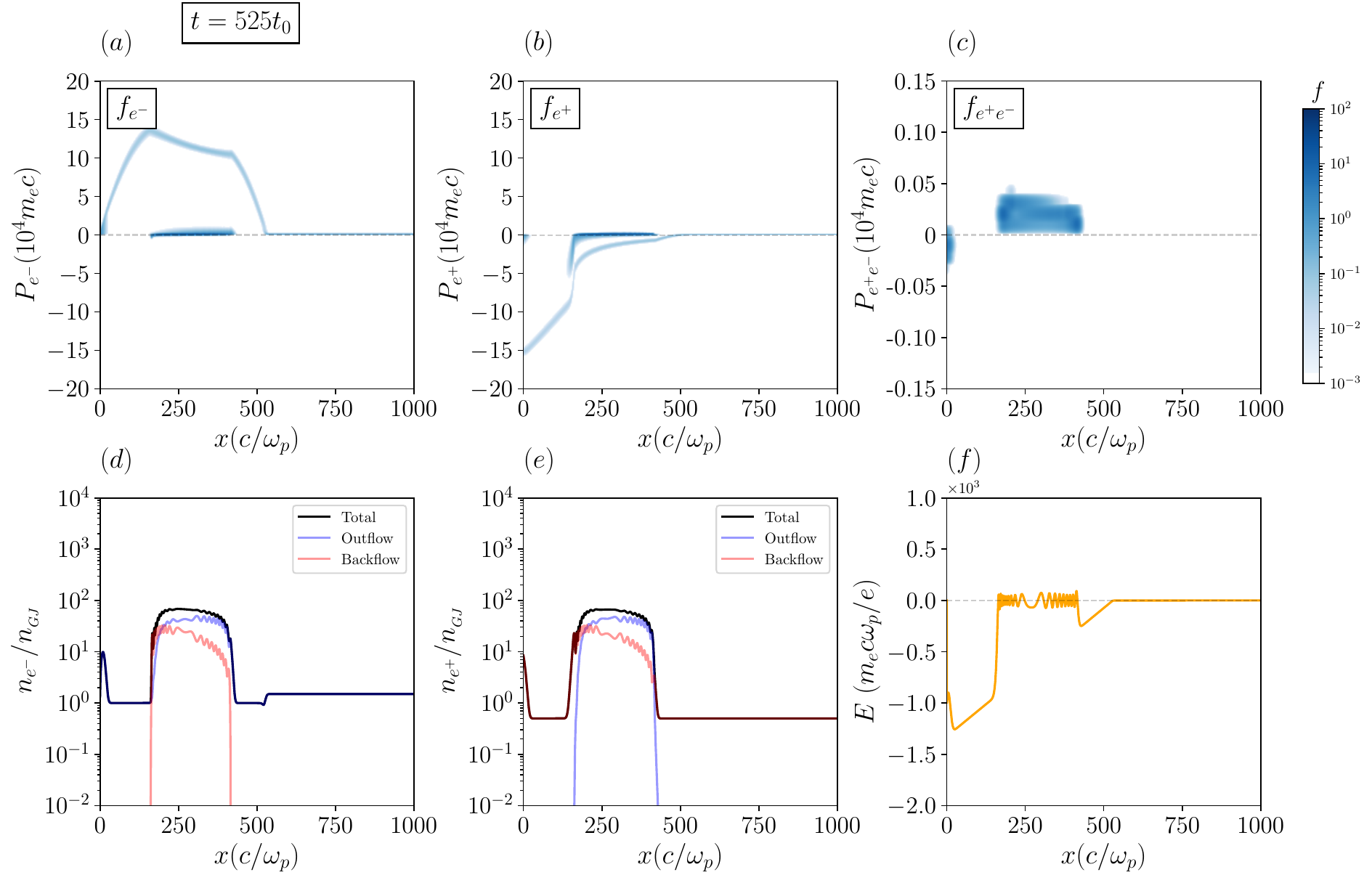}
    \caption{Snapshot of the simulation of the SCLF case at \( t = 525 t_0 \) (\(n_{\text{GJ}}=5n_0, \eta=2\)).In the SCLF model, the gap remains stationary during this stage, and pairs are primarily created near the center of the gap. Due to the presence of backflowing positrons, additional pair production occurs near the stellar surface. \\
    A full animation of the temporal evolution can be viewed at:\href{https://youtu.be/bwT6983U2_g}{SCLF-movie}}
    \label{fig:RS_snapshot}
\end{figure*}

Similar to the RS case, we consider the super-GJ scenario where $\eta > 1$. In our numerical tests, we find that in the super-GJ regime, the surface electric field always grows rapidly such that the resulting extraction flux quickly approaches the Goldreich–Julian limit \(n_{\rm GJ} c\), even when the tunable extraction fraction \(\zeta\) is set as low as 1\%. While all simulations presented in the following adopt \(\zeta = 1\%\), we retain the flexibility to vary \(\zeta\) in the code to facilitate future exploration of sub-GJ scenarios, where limited surface charge release may play a more significant role. It is known that for sub-GJ flow there exists a steady state with only modest particle acceleration and no pair production~\citep[e.g.][]{Chen2012}, therefore we focus on the regime where pair production is guaranteed to happen. We begin the simulation with the same initial conditions as in the RS model. As the initial electrons leave due to advection, the electric field near the pulsar surface grows at a rate given by $dE/dt = -(J_{\mathrm{ext}} - J_0),$ where \(J_{\mathrm{ext}}\) represents the extraction current density driven by the unscreened electric field at the surface, i.e., the space-charge-limited flow. In a fully unconstrained environment, $J_{\text{ext}}$ can easily reach $\rho_{\text{GJ}} c$, in which case 
\begin{equation}
    \frac{dE}{dt} = \frac{\eta - 1}{2\eta} J_0.
\end{equation}

\begin{figure*}[htbp]
    \centering
    \includegraphics[width=\textwidth]{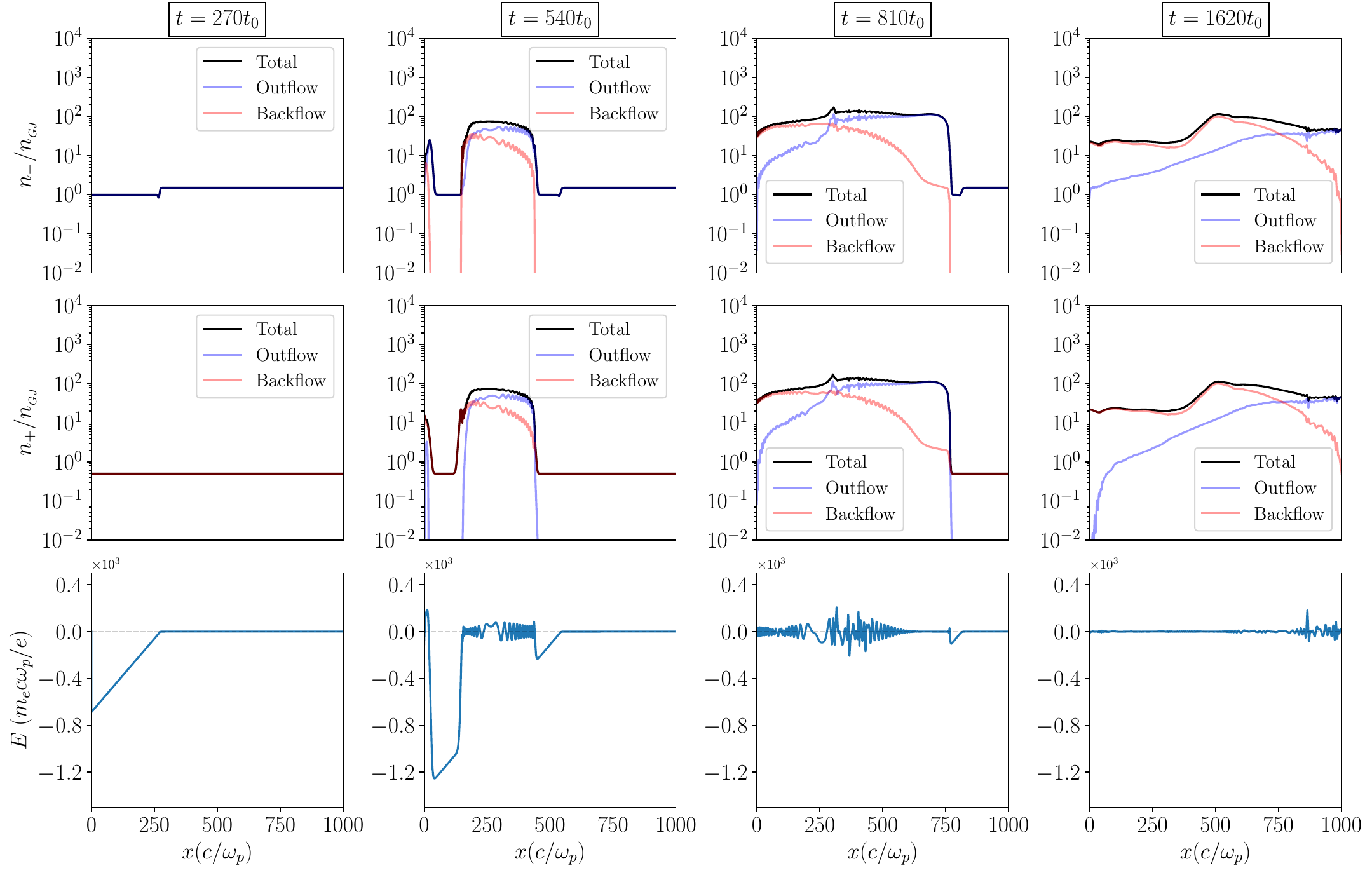}
    \caption{
    Time evolution of particle densities and electric field for an SCLF model discharge. Each column corresponds to a different snapshot in time, from left to right: \(t = 270t_0\), \(540t_0\), \(810t_0\), and \(1620t_0\).
    Top/Middle row: normalized electron/positron number density. 
    Bottom row: electric field profile. 
    Outflow and backflow components are shown separately in each density panel.
    The first column illustrates the initial gap formation phase.
    In the second column, a key difference between SCLF and RS models emerges: curvature photons emitted by primary electrons produce pairs near the center of the gap via magnetic conversion, while photons from backflowing positrons produce pairs later, near the stellar surface. The gap is screened on a timescale of approximately \(l_{\mathrm{gap}}/4c\).
    The third column shows that, unlike the RS model where pairs are produced in a narrow region at the left edge of the gap, the SCLF model generates pairs over a broad spatial extent. Moreover, since the accelerating field is screened in SCLF case, only primary charged particles with high momentum contribute to pair creation, resulting in a linearly increasing plasma density.
    The final column represents the relaxation phase.Due to the possibility of continuous electron extraction at the stellar surface in the SCLF model, the relaxation timescale can be extended.
    }
    \label{fig:SCLF_stages}
\end{figure*}

This corresponds to a relatively slow growth rate compared to the RS regime, especially when $\eta$ is not too much greater than 1. In this case, the particles being accelerated are the extracted electrons, whose momentum is given by 
\begin{equation}
    P_{e^-} =\int E\, dt = \frac{\eta - 1}{\eta} \left(t - \frac{x}{\beta}\right) \cdot \frac{x}{\beta} \, \Theta\left(\beta t - x\right) .
\end{equation}
Assuming that pair production is first triggered by CR photons emitted from electrons at the maximum momentum and following an analysis analogous to that for the RS model, the estimation of the length of the gap is given by
\begin{equation}
\begin{split}
    l_{\text{gap}}^{\text{SCLF}} &\simeq 2\times10^{4} R_{c,7}^{2/7} B_{12}^{-1/7} \chi^{1/7} n_{\text{GJ}}^{-3/7} (\eta - 1)^{-3/7}\ \text{cm}\\
    &\simeq 10^{4}  B_{12}^{-4/7} \chi^{1/7} P_1^{4/7} (\eta - 1)^{-3/7}\ \text{cm}.
\end{split}
\label{lgap_sclf}
\end{equation}
Under the SCLF model, electrons are expected to experience a time delay by a factor of 
\begin{equation}
    \frac{5}{6}\left[ \frac{\eta^{RS}+1}{\eta^{SCLF} -1}\frac{n^{RS}}{n^{SCLF}} \right]^{3/7}
\end{equation}
relative to the RS model before the onset of pair production. It is also important to note that the returning Gen~0 positrons are accelerated to substantial momenta while traversing the gap. However, the curvature photons they emit require a finite distance to propagate before the pair production. As a result, these photons typically convert into pairs near the pulsar surface, forming a secondary pair formation region. Moreover, in the limit where \( \eta \to 1 \), the entire simulation region begins to develop an electric field even before pair formation is initiated, and this estimate breaks down. Multi-dimensional effects likely need to be taken into account in this limit.

Apart from the factor of $\eta - 1$ in the gap size, the most fundamental difference between the RS and SCLF models arises during the second stage of the discharge process. It is at this stage that the presence or absence of an \emph{electron vacuum}, defined as a region with negligible electron density such that the local electric field remains unscreened, determines whether the gap develops into a propagating structure or remains localized, thereby shaping the overall discharge pattern in each model. In the SCLF model, the gap structure does not propagate outward as it does in the RS case.  Viewed in a reference frame co-moving with the gap at the speed of light, local electrons are promptly accelerated to the threshold momentum and immediately generate pairs, which then screen the electric field locally. As a result, the gap remains spatially confined and does not evolve into a propagating structure. In the SCLF model, the spatially confined gap is eventually screened out from the middle and disappears as the screening front propagates outward in both directions. This central screening is initiated by Gen~1 pairs generated by primary Gen~0 electrons. These Gen~1 particles encounter an already screened electric field and, as a result, do not reach the threshold to emit curvature photons or initiate further pair production. At the stellar surface, a latter-stage screening is provided by Gen~1 pairs produced by primary Gen~0 positrons that were previously accelerated through the gap. Together, these two-sided screening processes lead to the gradual vanishing of the gap structure. 

Due to the absence of an electron vacuum, the electric field in the SCLF model is efficiently screened once pair production is initiated. As a result, unlike the RS model, the SCLF cascade typically does not involve higher-generation particles (e.g., Gen~1+) in subsequent pair production. After the Gen~0 electrons are accelerated beyond the pair-production threshold and the electric field has been fully screened, they undergo only mild curvature radiation cooling. Before exiting the simulation domain, they continue to emit high-energy photons and produce electron-positron pairs, thereby sustaining the cascade for a finite duration even in the absence of an accelerating field. The growth of plasma density is therefore more linear in nature, reflecting the lack of sustained acceleration and feedback-driven multiplication.

The overall duration of the discharge cycle in the SCLF model can still be estimated using the same period formula derived for the RS model, i.e.\
\begin{equation}
    T_{SCLF} \approx \frac{1}{c} \left[ l_{\text{gap}}^{\text{SCLF}} + H \left( 1 + \frac{1}{2} \ln \left( \frac{\mathcal{N}_{\text{SCLF}}}{n_{\mathrm{GJ}}} \right) \right) \right].
\end{equation}
This estimate likely provides a lower bound for the actual cycle period in the SCLF model. This discrepancy arises from two key physical effects not accounted for in the estimate: First, the effective gap size in the SCLF model is longer due to the delayed onset of pair production. Second, during the late stages of the relaxation phase, additional electrons are continuously extracted and enter the simulation domain, effectively prolonging the decay timescale. As a result, the total cycle period in the SCLF model is expected to be longer than that in the RS case.

\section{Discussion}\label{sec:Discuss}

\begin{table*}[t]
\begin{tabular}{c c c c c c c c c}
\hline
Model & $n_{\text{GJ}}/n_0$ & $\eta$ & $H/x_0$ & $\mathcal{N}_{\text{peak}}$ & $l_{\text{gap}}^{\text{(Sim)}}/x_0$ & $l_{\text{gap}}^{\text{(Model)}}/x_0$ & $T^{\text{(Sim)}}/t_0$ & $T^{\text{(Model)}}/t_0$ \\
\hline\hline
$\; \star \mathrm{RS}_{1} \;$   & $\; 2 \;$ & $\; 1.5 \;$ & $\; 1000 \;$ & $\; \sim 3000 \;$ & $\; 509 \;$ & $\; 512 \;$ & $\; \sim 4500 \;$ & $\; \sim 4400 \;$ \\
$\; \mathrm{RS}_{2} \;$         & $\; 1 \;$ & $\; 1.5 \;$ & $\; 1000 \;$ & $\; \sim 3000 \;$ & $\; 686 \;$ & $\; 689 \;$ & $\;\sim 5100  \;$ & $\; \sim 5100 \;$ \\
$\; \mathrm{RS}_{3} \;$         & $\; 2 \;$ & $\; 2.0 \;$ & $\; 1000 \;$ & $\; \sim 4000  \;$ & $\; 470 \;$ & $\; 473 \;$ & $\; \sim 4800 \;$ & $\; \sim 4700 \;$ \\
$\; \mathrm{RS}_{4} \;$         & $\; 2 \;$ & $\; 1.5 \;$ & $\; 2000 \;$ & $\; \sim 4000
 \;$ & $\; 509 \;$ & $\; 512 \;$ & $\; \sim 9000 \;$ & $\; \sim 8900  \;$ \\
$\; \star \mathrm{SCLF}_{1} \;$ & $\; 5 \;$ & $\; 2.0 \;$ & $\; 1000 \;$ & $\; \sim 2800 \;$  & $\; 444 \;$ & $\; 426 \;$ & $\; \sim 4200 \;$ & $\; \sim 4000 \;$ \\
$\; \mathrm{SCLF}_{2} \;$ & $\; 4 \;$ & $\; 2.0 \;$ & $\; 1000 \;$ & $\; \sim 2000 \;$  & $\; 485 \;$ & $\; 469 \;$ & $\; \sim 4200 \;$ & $\; \sim 4000 \;$ \\
$\; \mathrm{SCLF}_{3} \;$ & $\; 5 \;$ & $\; 1.5 \;$ & $\; 1000 \;$ & $\;\sim 2000 
 \;$  & $\; 465 \;$ & $\; 575 \;$ & $\;\sim 4100\;$ & $\;\sim 4000  \;$ \\
\hline
\end{tabular}

\caption{Comparison between simulation results and model predictions for RS and SCLF cases. The fiducial runs analyzed in the main text are marked with $\star$. \(\mathcal{N}_{\text{peak}}\) corresponds to the highest number density reached during the simulation.}
\label{Comparison}

\end{table*}

Table~\ref{Comparison} summarizes the key simulation results and compares them with the predictions of our analytical model. The model-predicted gap length \(l_{\mathrm{gap}}^{(\mathrm{Model})}\) and period \(T^{(\mathrm{Model})}\) agree well with simulation in most RS cases, with relative errors typically below 2\%. This indicates that the model captures the scaling behavior of QED cascades over various physical conditions. The deviation in the SCLF period may be attributed to the more complex plasma loading process near the surface, which is not fully captured in the simplified model. For relatively large \(\eta\), the SCLF predictions remain in good agreement with the simulation. However, for smaller values of \(\eta\), such as \(\eta = 1.5\), the gap grows more slowly, potentially allowing pair production to proceed over an extended range, which introduces greater uncertainty in determining the effective gap length \( l_{\mathrm{gap}} \). We find that the analytical scaling for the SCLF model is most accurate for \(\eta \gtrsim 2\), and becomes somewhat less reliable as \(\eta\) approaches unity.

We also note a significant difference in the values of \( \mathcal{N}_{\text{peak}}/n_{\mathrm{GJ}} \), which reflects the distinct pair growth mechanisms in the two models. In the RS configuration, the presence of a vacuum for electrons enables newly created particles near the left edge of the gap to participate in the discharge process immediately. This leads to exponential pair multiplication, rapidly building up to multiplicities of \(10^3\)–\(10^4\) within the confined region. In contrast, the SCLF model lacks such an electron vacuum. As a result, only a few generations of pairs are needed to screen the accelerating field, and the original high-energy parent particles dominate further pair production. This leads to a much slower, essentially linear, growth in plasma density. Under comparable parameters, the local multiplicity in the SCLF model may reach only about 20\% of that in the RS model, providing a useful estimate for the expected range of pair multiplicities in realistic pulsar environments, which likely lie between these two extremes. After the peak-density plasma blob exits the simulation domain, the residual plasma density \(\mathcal{N}\) left behind is typically only 20\%--30\% of the peak value. This remnant density provides a practical estimate of the relaxation-stage duration, during which the system evolves toward the background state before the next discharge is triggered. 

\subsection{Energy Budget and Surface Heating by Returning Particles}

To estimate the total energy supplied by the magnetic field, we start from the evolution of the electric field energy, defined as
\begin{equation}
    U_E(t) = \int_0^H \frac{1}{2} E^2(x, t) \, dx.
\end{equation}
Taking the time derivative and substituting the field equation \ref{revised_maxwell} we obtain
\begin{equation}
    \frac{dU_E}{dt} = \int_0^H (J_0 - J) E \, dx.
\end{equation}
This expression shows that the energy is injected by the external current \( J_0 \), while the local current \( J \) dissipates it. The total injected energy from the magnetic field over a full discharge cycle is therefore given by
\begin{equation}
    \int_{t}^{t+T} \int_0^H J_0 E \, dx\, dt,
\end{equation}
which sets the upper limit on the energy available to accelerate particles and heat the stellar surface. We note that the dominant phase of magnetic energy injection differs in timing between the two models. In the RS configuration, most of the energy is injected during the outward propagation of the gap structure, as it leaves the simulation domain. In the SCLF model, by contrast, energy injection primarily occurs shortly after gap formation, when the accelerating electric field is rapidly screened by newly created pairs. Neglecting field oscillations caused by plasma propagation, we can estimate the total magnetic energy injected into the domain over one discharge cycle. For the RS model, this energy is primarily delivered during the outward motion of the gap and is approximately given by
\begin{equation}
    U_{\text{in}}^{\text{RS}} \sim l_{\text{gap}}^{\text{RS}} \left[ \frac{1}{2}  \frac{\eta + 1}{2\eta}  J_0 \frac{l_{\text{gap}}^{\text{RS}}}{c} \right]  J_0  \frac{H}{c} ,
\end{equation}
where the terms in brackets represent the mean electric field in the gap, multiplied by the effective duration of energy injection. For the SCLF model, the corresponding estimate is
\begin{equation}
    U_{\text{in}}^{\text{SCLF}} \sim l_{\text{gap}}^{\text{SCLF}} \left[ \frac{1}{2}  \frac{\eta - 1}{2\eta}  J_0  \frac{l_{\text{gap}}^{\text{SCLF}}}{c} \right]  J_0\frac{l_{\text{gap}}^{\text{SCLF}}}{4c} .
\end{equation}
Assuming that the discharge cycle has a characteristic duration \( T \sim 4H/c \), we can estimate the average magnetic energy injection rate into the domain. For the RS model, the estimated magnetic energy injection rate is on the order of
\begin{equation}
\begin{split}
    \mathcal{P}^{\text{RS}}&\sim10^{32}R_{c,7}^{4/7} B_{12}^{-2/7} \chi^{2/7} n_{\text{GJ}}^{8/7} (\eta + 1)^{1/7}\eta\ \text{erg/s}\\
    &\sim 10^{32} B_{12}^{6/7} \chi^{2/7} P_1^{-6/7} (\eta + 1)^{1/7}\eta\ \text{erg/s}.
\end{split}
\label{Heatingrate}
\end{equation}

For identical parameters, the energy injection rate in the SCLF model is suppressed relative to the RS model by a factor of
\begin{equation}
    \frac{\mathcal{P}^{\text{SCLF}}}{\mathcal{P}^{\text{RS}}} \sim \left( \frac{\eta - 1}{\eta + 1} \right)^{1/7} \frac{l_{\text{gap}}^{\text{SCLF}}}{4H}.
\end{equation}
The magnetic energy injection rate calculated from the simulations agrees with the model prediction within 10\% across all cases, indicating that the model captures the essential scaling of the electromagnetic energy budget. As discussed in the previous section, this estimate is only valid when \( \eta \) is significantly different from unity. In the limit \( \eta \to 1 \), the accelerating electric field extends throughout the domain, and the notion of a well-defined gap structure breaks down.

A substantial fraction of this injected energy is carried by returning particles, which deposit their kinetic energy onto the stellar surface and contribute to polar cap heating, i.e., the corresponding heating rate \(\mathcal{Q} \propto \mathcal{P}.\) We computed the fraction of this injected magnetic energy that is carried by returning particles and deposited onto the stellar surface. This return current heating accounts for approximately 30--50\% of the total injected energy, depending on the model. In general, SCLF cases exhibit a slightly higher return fraction than RS cases, with some SCLF configurations reaching or slightly exceeding 50\%. The resulting heating rate, on the order of \(\sim 10^{32}~\mathrm{erg\cdot s^{-1}}\), is consistent with previous theoretical predictions \citep{sznajder2020}.

\subsection{The Spatiotemporal and Spectral Structures of the Electric Field Energy} \label{Spectrum}

\begin{figure}[htbp]
    \centering
    \includegraphics[width=0.5\textwidth]{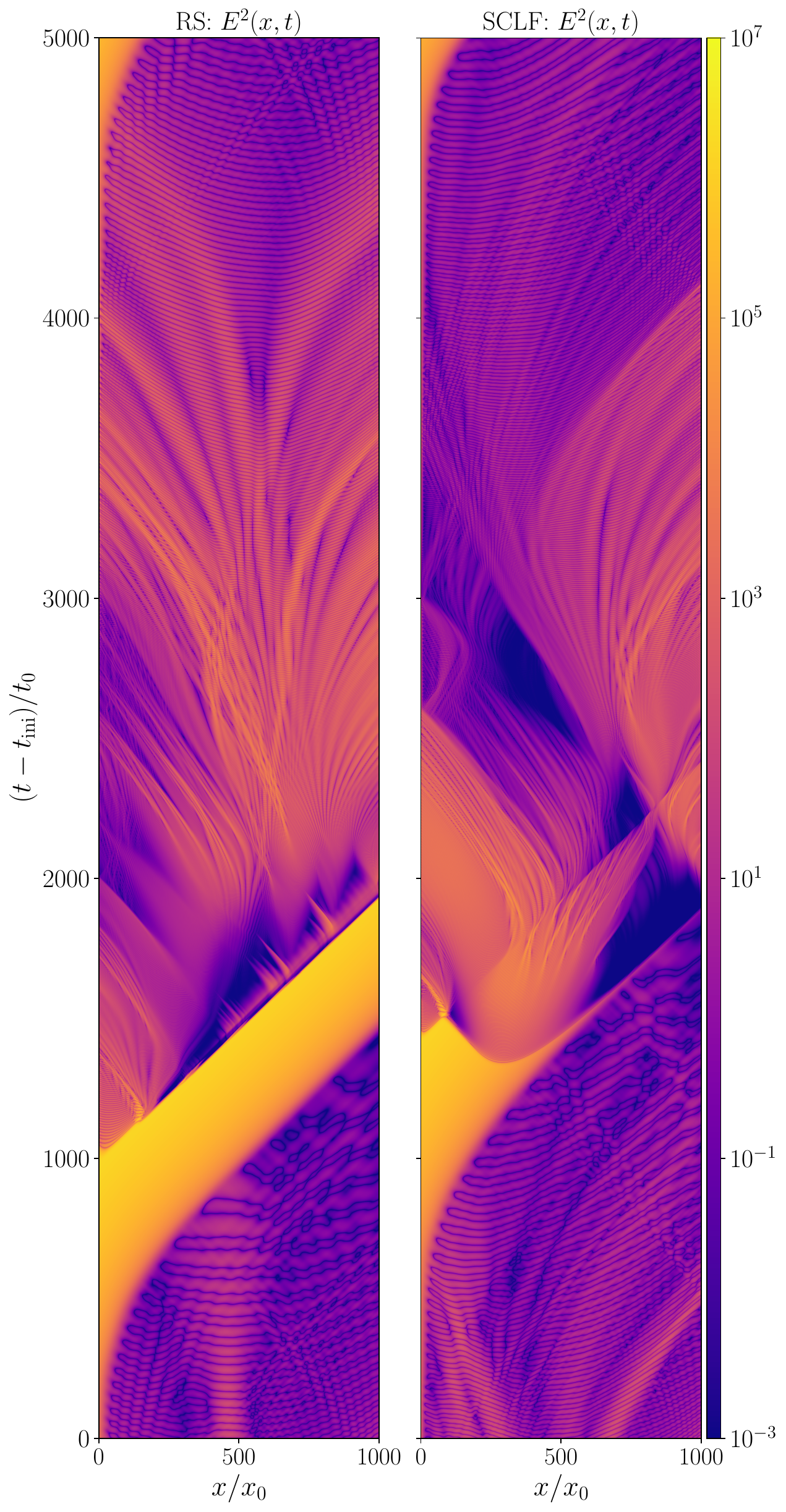}
    \caption{
    Spatiotemporal evolution of electric field energy density $E^2(x,t)$ for two cascade models.
    The left panel shows the Ruderman–Sutherland case, and the right panel shows the Space-Charge-Limited Flow case. The vertical axis is rescaled as $(t - t_{\mathrm{ini}})/t_0$ with each plot using its own $t_{\mathrm{ini}}$ offset. Strong spatiotemporal oscillations and mode structures are clearly visible in both models. A noteworthy difference is that the RS solution maintains an outward-propagating gap structure, whereas in the SCLF model, the gap mostly remains confined to the pulsar surface.
    }
    \label{fig:E2xt}
\end{figure}

\begin{figure*}[htbp]
    \centering
    \includegraphics[width=1\textwidth]{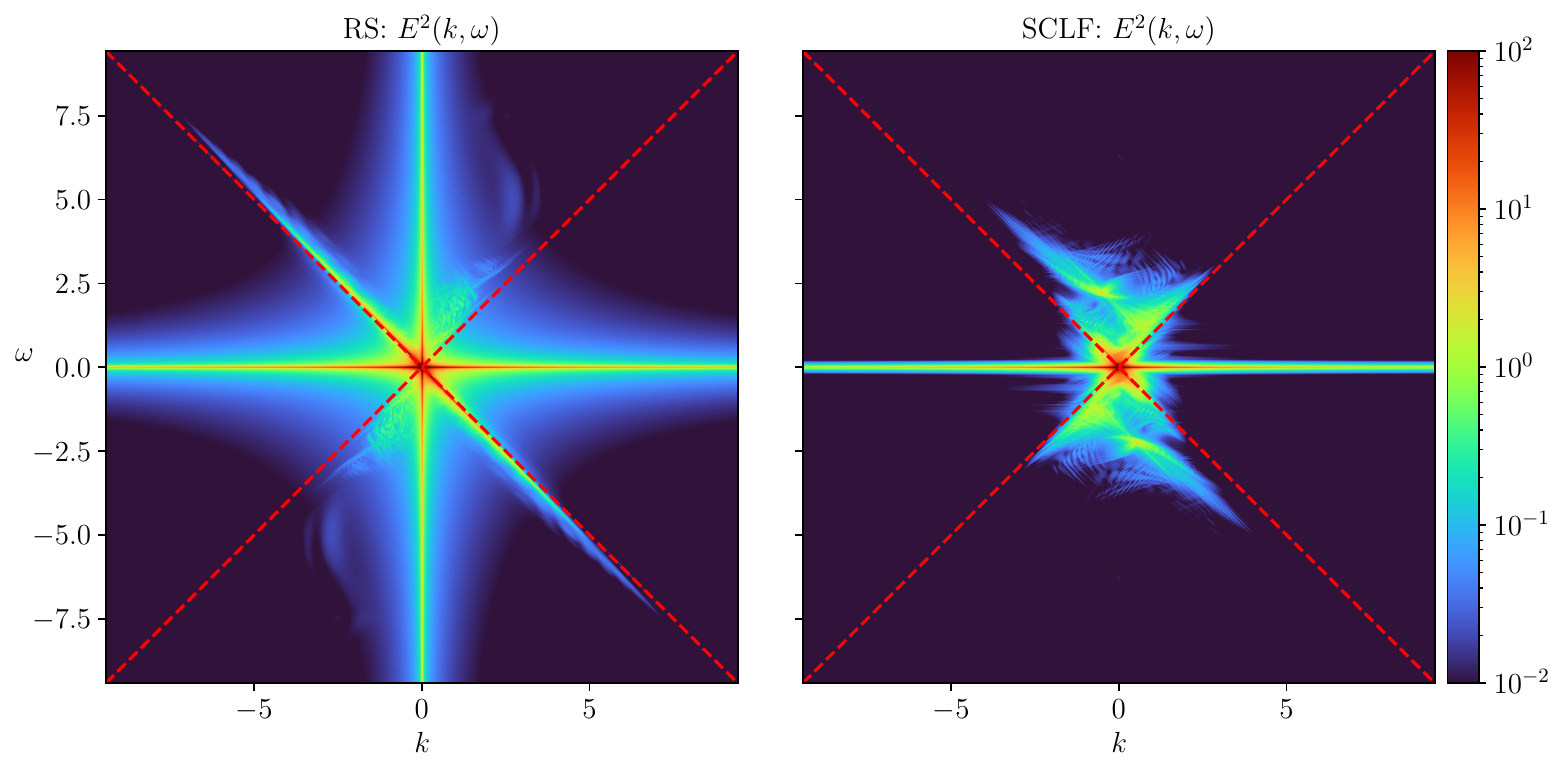}
    \caption{
    Frequency–wavenumber spectrum of the electric field energy density $E^2(k,\omega)$ for the RS (left) and SCLF (right) models. The red dashed lines indicate $\omega = \pm ck$, corresponding to the light cone.
    The wave components near $\omega = \pm ck$ are likely driven by ultra-relativistic plasmas. Moreover, the spectrum shows the presence of complex superluminal modes, generally believed to be intricate mixtures of superluminal O-modes.
    }
    \label{fig:E2kw}
\end{figure*}

The spatiotemporal and spectral structures of the electric field energy \(E^2\) in both models are shown in Figure~\ref{fig:E2xt} and Figure~\ref{fig:E2kw}. The large, coherent structures in Figure~\ref{fig:E2xt} starting around time ${\sim}1000$ corresponds to the gap. The figure also clearly shows strands of characteristics of plasma waves after the gap passes through the domain or is screened. These strands show very rapid oscillations initially (as can be seen by zooming in on the figure). The oscillation frequency becomes much lower at later times, when the plasma density drops closer to $\rho_\mathrm{GJ}$. The spectral plot, Figure~\ref{fig:E2kw}, illustrates the distribution of wave energy across wave number \(k\) and angular frequency \(\omega\), thereby highlighting the dominant propagating modes. In both plots, diagonal features aligned with the trajectories $x = ct$ and \(\omega = \pm ck\) are clearly visible, indicating the presence of electromagnetic modes propagating at or near the speed of light. These structures are likely associated with electric field oscillations driven by charge carriers traveling at relativistic velocities. 

In the RS case, the spectrum exhibits a distinct vertical feature at \(k = 0\), indicating the presence of spatially uniform field components. This feature likely arises from the uniform pair production induced by the spatially propagating gap, which imprints a domain-wide structure onto the electric field. Additionally, the spectrum shows a pronounced asymmetry in \(\omega\), with enhanced power along the \(\omega = -ck\) branch. This suggests that the backflow is primarily composed of super-relativistic, beam-driven waves. In contrast, the outflow signal appears above the \(\omega = ck\) line but is more broadly distributed in \((k, \omega)\), possibly indicating a complex mixture of wave modes. This spectral asymmetry may help explain why the backflow heating rate typically accounts for only about 40\% of the total injection power.

In the SCLF case, the absence of a significant \(k = 0\) component indicates that the excited modes lack large-scale spatial uniformity, which is consistent with the fact that the gap develops and is screened locally near the stellar surface. The spectral energy is broadly distributed in regions where \(|\omega/k| \geq c\), which may suggest the presence of superluminal electromagnetic modes. The relatively symmetric distribution in \(k\) reflects bidirectional wave propagation and may help explain why the backflow heating rate typically accounts for approximately 50\% of the total injection power.

However, due to the 1D nature of the simulation, the mode identification remains ambiguous. \citet{Philippov2020} interpreted similar wave structures observed in 2D simulations as superluminal O-modes, and suggested that the corresponding 1D features could be projections of such modes. This interpretation, however, may be questionable. In particular, since our simulation does not evolve the full electromagnetic fields, the superluminal O-modes \emph{should not have been captured in the first place}. Additionally, according to \citet{Rafat2019b}, the superluminal component observed in 1D may instead correspond to a Langmuir-type wave that appears superluminal due to a Lorentz boost. In this case, the 1D structure would not be a projected O-mode but rather a fundamentally different mode that potentially contaminates the interpretation of wave dynamics in higher dimensions. In a multi-dimensional system, a more thorough analysis of the wave modes is required to determine the nature of these superluminal waves.
Unfortunately, the ambiguity cannot be resolved in the present 1D setup. We may explore this question in more detail in future studies.

\subsection{Scope of Quantitative Estimates in Millisecond Pulsars}

Millisecond pulsars (MSPs) represent a distinct subpopulation of neutron stars characterized by exceptionally short rotational periods (typically $P \lesssim 10$~ms) and remarkably low period derivatives. In contrast, canonical pulsars generally exhibit spin periods on the order of $0.1$--$1$~s and magnetic field strengths of ${\sim}10^{12}$~G. This dichotomy stems from their differing evolutionary paths: canonical pulsars are born in core-collapse supernovae with strong magnetic fields and undergo rapid magnetic braking, whereas MSPs are thought to be ``recycled'' via accretion-induced spin-up in low-mass X-ray binaries (LMXBs) \citep{Bhattacharya1991}.  Quantitatively, these distinctions manifest in key observable and derived parameters such as surface dipolar magnetic field strength, spin-down luminosity, and characteristic age. MSPs typically exhibit $B \sim 10^8$--$10^9$~G, $\dot{P} \sim 10^{-20}$--$10^{-19}$, and characteristic ages $\tau_c \gtrsim 10^9$~yr, placing them among the oldest known neutron stars in the Galaxy. \citep{Lorimer2005}

The striking difference in the characteristic parameters of MSPs and canonical pulsars—often spanning several orders of magnitude—motivates a careful assessment of the general applicability of 
our gap models presented in Section~\ref{sec:sim}.
The electromagnetic power, plasma densities, or gap potentials were originally derived with young pulsars in mind, and it is not immediately clear whether these relations remain valid for MSPs, whose magnetospheric and plasma environments are markedly different. To address this, we perform quantitative comparisons using representative parameters, the spin period $P$ and surface magnetic field $B$, for both classes of pulsars. By applying the same estimation formulae introduced in this work to both canonical pulsars and MSPs, we evaluate their scaling behaviors and assess the robustness of our model across the pulsar population.

Consider a millisecond pulsar (labeled MSP) with period \( P = 5\,\mathrm{ms} \) and surface magnetic field \( B = 10^8\,\mathrm{G} \), we compare it with a canonical pulsar (labeled CP) with period $P = 1\,\mathrm{s}$ and surface magnetic field $B = 10^{12}\,\mathrm{G}$.

Substituting this into Equation~\eqref{lgap} yields a gap length of:
\begin{equation}
    l_{\mathrm{gap}}^\mathrm{MSP}\sim \left(\frac{B_\mathrm{MSP}}{B_\mathrm{CP}}\right)^{-4/7}\left(\frac{P_\mathrm{MSP}}{P_\mathrm{CP}}\right)^{4/7}l^\mathrm{CP}_{\mathrm{gap}} \approx 10\ l^\mathrm{CP}_{\mathrm{gap}}.
\end{equation}
Since the gap size in the RS model and the SCLF model mostly differ by a factor related to $\eta$, they follow the same scaling with respect to $P$ and $B$.

On the other hand, the polar cap radius $r_\mathrm{pc}$ increases significantly for millisecond pulsars:
\begin{equation}
    r_\mathrm{pc}^\mathrm{MSP}\sim \left(\frac{P_\mathrm{MSP}}{P_\mathrm{CP}}\right)^{-1/2}r_\mathrm{pc}^\mathrm{CP}\approx 14\,r_\mathrm{pc}^\mathrm{CP},
\end{equation}
therefore the ratio $l_\mathrm{gap}/r_\mathrm{pc}$ remains roughly the same despite the significant order-of-magnitude change in parameters. As a result, our 1D model should still hold well in the case of millisecond pulsars. Additionally, even though the curvature radius of magnetic field lines $R_c$ is significantly lower in the MSP case, it remains much larger than the gap size $l_\mathrm{gap}$, therefore there is no significant deviation from our photon free path approximation $\ell_\mathrm{ph} \sim 2R_c/\epsilon_\gamma$.

\section{Conclusions}

In this work, we investigated the nonlinear evolution of vacuum gaps and electric field structures in pulsar polar cap environments, focusing on the Ruderman--Sutherland and Space-Charge-Limited Flow models. Using a custom-developed GPU-accelerated relativistic Vlasov code named \texttt{PRVMs}, we performed one-dimensional kinetic simulations that resolve the fine-scale dynamics of charge-separated plasma and gap oscillations. The code enables high-resolution modeling of pair creation, wave propagation, and field feedback in regimes inaccessible to fluid or hybrid approaches.

Our simulations reveal distinct behaviors between the two models: the RS configuration supports a persistent, outward-propagating gap structure, while in the SCLF model the gap remains confined to the stellar surface. We derived approximate scaling relations for the gap width and energy injection rate, and validated them against numerical results. These relations provide physical insight into how pair multiplicity, magnetic field strength, and curvature geometry affect the local gap dynamics. Our model largely agrees with what was proposed by~\citet{2015ApJ...810..144T} and~\citet{2019ApJ...871...12T}, although we focus more on the in-gap energy dissipation, and the comparison of RS and SCLF regimes.


Future work will extend this study in multiple directions. In particular, we aim to relax the one-dimensional assumption and include transverse transport and realistic magnetospheric geometries. The modular structure and GPU scalability of \texttt{PRVMs} make it well-suited for such high-dimensional studies. Ultimately, these improvements will allow us to self-consistently capture the interplay between gap electrodynamics and collective plasma processes that may underlie pulsar radio emission. In addition, resolving the spatial and temporal structure of particle precipitation will enable predictions of surface heating patterns, potentially linking our models to X-ray observations of hot spot morphology.

\begin{acknowledgments}
We thank Anatoly Spitkovsky and Yajie Yuan for helpful discussions. DY and AC acknowledge support from NSF grants DMS-2235457 and
AST-2308111. AC acknowledges additional support from NASA grant
80NSSC24K1095. The authors acknowledge the Research Infrastructure Services (RIS) group at Washington University in St. Louis for providing computaional resources and services needed to generate the research results delivered within this paper.
\end{acknowledgments}

\bibliographystyle{aasjournal}
\bibliography{mybib} 
\end{document}